\documentclass[num-refs]{wiley-article}

\usepackage{hyperref}
\usepackage[utf8]{inputenc}
\usepackage{dirtytalk}
\usepackage{tcolorbox}
\usepackage{multirow}
\usepackage{framed}

\definecolor{formalshade}{rgb}{0.97,0.97,0.97}
\definecolor{darkblue}{rgb}{0.90,0.90,0.90}

\newenvironment{formal}{%
  \MakeFramed{\advance\hsize-\width\FrameRestore}%
  \noindent\hspace{-4.55pt}
  \begin{adjustwidth}{}{7pt}%
  \vspace{2pt}\vspace{2pt}%
}
{%
  \vspace{2pt}\end{adjustwidth}\endMakeFramed%
}

\usepackage{fontawesome}
\usepackage{siunitx}

\title{A Qualitative Study of Architectural Design Issues in DevOps}

\author[1]{Mojtaba Shahin*}
\author[2]{Ali Rezaei Nasab}
\author[3]{Muhammad Ali Babar}

\affil[1]{Faculty of Information Technology, Monash University, Melbourne, Victoria, Australia, {mojtaba.shahin@monash.edu}}
\affil[2]{Department of Engineering, Computer Science and Information Technology, Shiraz University, Shiraz, Iran, {rezaei.ali.nasab@gmail.com}}

\affil[3]{School of Computer Science, University of Adelaide, Adelaide, South Australia, Australia, {ali.babar@adelaide.edu.au}}

\corraddress{*Faculty of Information Technology, Monash University, Melbourne, Victoria, Australia}
\corremail{mojtaba.shahin$@$monash.edu}

\runningauthor{Shahin et al.}

\begin{document}

\begin{frontmatter}
\maketitle

\begin{abstract}
Software architecture is critical in succeeding with DevOps. However, designing software architectures that enable and support DevOps (DevOps-driven software architectures) is a challenge for organizations. We assert that one of the essential steps towards characterizing DevOps-driven architectures is to understand architectural design issues raised in DevOps. At the same time, some of the architectural issues that emerge in the DevOps context (and their corresponding architectural practices or tactics) may stem from the context (i.e., domain) and characteristics of software organizations. To this end, we conducted a mixed-methods study that consists of a qualitative case study of two teams in a company during their DevOps transformation and a content analysis of Stack Overflow and DevOps Stack Exchange posts to understand architectural design issues in DevOps. Our study found eight specific and contextual architectural design issues faced by the two teams and classified architectural design issues discussed in Stack Overflow and DevOps Stack Exchange into 11 groups. Our aggregated results reveal that the main characteristics of DevOps-driven architectures are: being loosely coupled and prioritizing deployability, testability, supportability, and modifiability over other quality attributes. Finally, we discuss some concrete implications for research and practice.

\end{abstract}
\end{frontmatter}

\section{Introduction}\label{secintro}

The emergence of Development and Operations (DevOps) paradigm as a promising approach to build and release software at an accelerated pace has stimulated widespread industrial interests \cite{forsgren20172017}. However, establishing DevOps culture (e.g., shared responsibility) and implementing its practices such as continuous delivery and deployment require new organizational capabilities and innovative techniques and tools for some, if not all, software engineering activities \cite{shahin2017continuous,bass2015devops,leite2019survey}.
There has been an increasing amount of literature on DevOps, which mostly deals with integrating security \cite{rahman2016software,jaatun2017devops}, enhancing test and deployment automation \cite{makinen2016improving,kang2016container,wettinger2016streamlining}, and improving performance \cite{van2017report} in DevOps pipelines. Other studies report the challenges (e.g., limited visibility on customer environments) that organizations encountered in DevOps adoption and the practices (e.g., test automation) employed to address those challenges \cite{lwakatare2019devops,luz2019adopting,erich2017qualitative,lwakatare2016towards}. Research also discussed the required changes in organizational structures and developers’ skills and responsibilities for adopting DevOps \cite{shahin2017adopting,nybom2016impact,skelton2019team,leite2020organization}.

It is becoming important to understand how an application should be (re-) architected to support DevOps \cite{forsgren20172017,bass2015devops}. Software architecture is expected to be the keystone for reaching the highest level of DevOps success \cite{shahin2017continuous,leite2019survey}. Most of the reported research on software architecture and DevOps relatedness studied software architecture in continuous delivery and deployment as a key practice of DevOps \cite{shahin2019empirical,bellomo2014toward,chen2015towards} or only focused on the microservices architecture style as a promising style to enable DevOps \cite{balalaie2016microservices,callanan2016devops,chen2018microservices,zdun2019emerging,waseem2020systematic}. For example, Shahin et al. \cite{shahin2019empirical} focused on the perception of practitioners from different organizations around the world on how software architecture being impacted by or impacting continuous delivery and deployment, resulting in a conceptual framework to support the process of (re-) architecting for continuous delivery and deployment. Balalaie et al. \cite{balalaie2016microservices} showed that the microservices architecture style facilitates the adoption of DevOps.

While DevOps, continuous delivery and deployment, and microservices share common characteristics (e.g., automating repetitive tasks) and reinforce each other \cite{Exploring,SchmidtDevOps}, organizations may adopt only one of these practices to achieve their business goals, e.g., delivering quality software with a shorter time more reliably \cite{laukkarinen2017devops}. The constraints imposed by the domains of organizations may prevent some of them from pushing software changes to production continuously (continuous delivery and deployment practice) \cite{lwakatare2016towards,shahin2017beyond}, while they can still implement other DevOps practices (e.g., Infrastructure as Code). Also, the microservices architecture style is still not easily used by many software organizations due to its challenging nature \cite{taibi2017microservices, whenuseMicroservices}. It is argued that the microservices architecture style needs highly skilled developers and requires a deep understanding of new technologies \cite{taibi2017microservices, leite2019survey}. Hence, organizations might use different approaches to (re-) design software architectures to enable or support DevOps.

There is still little research that has systematically studied software architectures that enable and support DevOps (here referred to as DevOps-driven software architectures) \cite{leite2019survey}. We believe that one of the essential steps towards characterizing DevOps-driven software architectures is to understand and analyze the types of architectural design issues that software practitioners face in the context of DevOps. At the same time, some of the architectural issues that emerge in the DevOps context (and their corresponding architectural practices or tactics) may be tightly associated with organizational context (i.e., domain) \cite{zhu2016devops}.

To understand the types of architectural design issues in DevOps, we carried out a mixed-methods empirical study consisting of a qualitative case study with two teams in a company, complemented with a content analysis of Stack Overflow and DevOps Stack Exchange posts. The qualitative case study aims to understand the specific and contextual architectural design issues faced by the two teams during DevOps transformation and the architectural decisions made to address each design issue. The content analysis of Stack Overflow and DevOps Stack Exchange posts concerning DevOps and architecture enables us to understand architectural design issues in DevOps from a more general perspective.

Our study leads to the following findings and contributions:

\begin{itemize}
   \item We identified eight specific and contextual architectural design issues that two teams from a company faced in DevOps. Although sometimes the two teams used different practices or made different decisions to address these design issues, the practices and decisions mainly aimed to improve deployability, testability, supportability, and modifiability.
    
    \item We identified 11 architectural design issue groups in the context of DevOps in Stack Overflow and DevOps Stack Exchange. Among these 11 architectural design issue groups, \say{Configuration} and \say{The complexity of (Micro) Services at the Design Level} are the top two frequently discussed architectural design issues in the context of DevOps. \say{Availability and Scalability of Services} and \say{Test} are identified as the most popular architectural design issues. \say{Deployment}, \say{Security}, and \say{Test} are flagged as the most challenging architectural design issues.
    \item All the eight architectural design issues from the two teams, except one, can be linked to one or more architectural design issue groups identified from Stack Overflow and DevOps Stack Exchange. Further to this, we found design issues related to security, database, and container discovered by the content analysis study have no corresponding issue in the case study.
    \item The main characteristics of DevOps-driven architectures are: being loosely coupled and prioritizing deployability, testability, supportability, and modifiability over other quality attributes.
    \item \say{Configuration} is the most discussed issue in Stack Overflow and DevOps Exchange, and the members of TeamA and TeamB faced challenges in doing operations tasks. Hence, we argue that operations specialists should be part of the software development teams to perform the operations tasks that require advanced expertise.
    \item We argue that software organizations and practitioners should significantly invest in automated testing (automating tests that occur during the last stages of DevOps pipelines) if they want to release software changes continuously.
\end{itemize}

We have previously reported the design and findings of the qualitative case study (On the Role of Software Architecture in DevOps Transformation: An Industrial Case Study) at the ICSSP conference in 2020 \cite{shahin2020}. We extend the previously published work with a content analysis of 100 posts from Stack Overflow and DevOps Stack Exchange. This paper also compares and contrasts the findings of the qualitative case study and the content analysis study.

\textbf{Paper Organization}: Section \ref{secRelatedWork} summarizes the related work. Section \ref{secresearch} outlines our research method. We present our findings in Section \ref{secfindings}. Section \ref{discussionSec} reflects on our findings, and we then discuss the threats to our study’s validation in Section \ref{secThreats}. Finally, we close the paper in Section \ref{secConclusion}.

\section{Related Work}\label{secRelatedWork}
Existing research investigated the role of software architecture in continuous integration and continuous delivery and deployment as two practices of DevOps or leveraged microservices architecture to support DevOps.

Over the last eight years, Puppet \footnote{https://puppet.com/}  has annually released non-peer reviewed reports to study the current state of DevOps in practice \cite{forsgren20172017}. In 2017 \cite{forsgren20172017}, the role of software architecture in DevOps was deeply examined to investigate how application architecture, and the structure of the teams that work on, impact the delivery capability of organizations. The main finding of this report reads, \say{loosely coupled architectures and teams are the strongest predictors of continuous delivery}. Surprisingly, it was found that many so-called service-oriented architectures (e.g., microservices) in practice may prevent testing and deploying services independently from each other. Subsequently, it can negatively influence teams to develop and deliver software.

Shahin et al. \cite{shahin2019empirical} have conducted a mixed-methods study to explore how software architecture is being impacted by or is impacting continuous delivery and deployment. They present a conceptual framework to support (re-) architecting for continuous delivery and deployment. The work by Mårtensson et al. \cite{maartensson2017continuous} identified 12 enabling factors (e.g., \say{work breakdown} and \say{test before commit}) impacting the capability of the developers when practicing continuous integration. The study \cite{maartensson2017continuous} argues that some of these factors (e.g., work breakdown) can be limited by the architecture of a system. Shahin et al. \cite{shahin2019empirical}, Chen \cite{chen2015towards}, and Chen \cite{chen2018microservices} argue that a set of quality attributes such as deployability, security, modifiability, and monitorability, require more attention when designing architectures in the context of continuous delivery and deployment. Shahin et al. \cite{shahin2019empirical} also found that a lack of explicit control on reusability poses challenges to practicing continuous delivery and deployment. The studies \cite{shahin2019empirical, schermann2016empirical} found that monoliths are the main source of pain to practice continuous delivery and deployment in the industry. In a retrospective study on three projects adopting continuous integration and delivery, Bellomo et al. \cite{bellomo2014toward} revealed that the architectural decisions made in those projects played a significant role in achieving the desired state of deployment (i.e., deployability). Di Nitto et al. \cite{di2016software} outlined architecturally significant stakeholders (e.g., infrastructure provider) and their concerns (e.g., monitoring) in DevOps scenarios. Then a framework called SQUID was built, which aims at supporting the documentation of DevOps-driven software architectures and their quality properties. Openja et al. \cite{Moses2020Analysis} studied modern release engineering topics and their difficulty in Stack Overflow using the Latent Dirichlet allocation (LDA). Among 38 identified release engineering topics grouped in 16 categories, questions in the \say{Continuous Integration/Continuous Deployment} category were the most prevalent. Further to this, \say{Mobile App Debug \& Deployment} and \say{Continuous Deployment} topics were the most complicated.

Another line of research has investigated microservices architecture in DevOps. Waseem et al. \cite{waseem2020systematic} conducted a systematic mapping study on 57 studies to characterize microservices architecture in the context of DevOps. They found three key research topics on microservices architecture in DevOps: \say{microservices development and operations in DevOps}, \say{approaches and tools support for microservices systems in DevOps}, and \say{migrations from monoliths to microservices architectures in DevOps}. Based on an experience report, Balalaie et al. \cite{balalaie2016microservices} presented the architectural patterns (e.g., change code dependency to service call) and technology decisions (e.g., using containerization to support continuous delivery) employed by a case company to re-architect a monolithic architecture into microservices in the context of DevOps. They also reported that a microservices-based architecture might pose some challenges, such as bringing higher complexity. Callanan and Spillane \cite{callanan2016devops} discuss that developing a standard release path and implementing independently releasable microservices through building backward compatibility with each release were the main tactics leveraged by their respective company to smooth DevOps transformation. These tactics also significantly reduced delays in the deployment pipeline. 

Our study is different from the works mentioned above. (1) Our work provides an in-depth analysis of how two teams in a company architect their systems during DevOps transformation. Due to certain circumstances, in contrast to \cite{shahin2019empirical, balalaie2016microservices}, these two teams have not adopted the microservices architecture style or continuous delivery and deployment in their DevOps transformation. (2) Our paper investigates concrete supporting technologies to support architectural decisions in DevOps, which has not been explored in previous works. (3) Our work is the first (to the best of our knowledge) to explore architectural design issues reported in question and answer websites (e.g., Stack Overflow and DevOps Stack Exchange).

\section{Research Design}\label{secresearch}
Our goal is to understand architectural design issues in DevOps. We decompose this goal into the following research questions.

\textbf{RQ1. What architectural design issues are raised during the transformation of a company to DevOps?}

\underline{\textbf{Rationale}}. Software organizations vary in their development methods, resources (e.g., infrastructures, team members), structures, domains, cultures, etc. \cite{capilla201610}. These differences lead to software organizations face different (and sometimes unique) challenges when developing software and consequently employ different behavior to overcome these challenges. Hence, this research question aims to understand architectural design issues and their corresponding architectural decisions in the context of a specific company during the DevOps transformation.

\textbf{RQ2. What architectural design issues in DevOps are raised in question and answer websites?}

\underline{\textbf{Rationale}}. Investigating architectural design issues in the context of DevOps from only a company's perspective can provide useful insights. However, there is a chance to miss some other important architectural design issues. A wide range of software practitioners (e.g., apps developers, designers) with varied expertise and seniority level from different countries, cultures, and companies use Stack Overflow and DevOps Stack Exchange websites to share and leverage knowledge on a variety of topics (e.g., architectural knowledge) \cite{oliveira2018exchange, vadlamani2020studying, soliman2016architectural}. Hence, investigating architectural design issues in the DevOps context reported in Stack Overflow and DevOps Stack Exchange can complement RQ1 and enable us to gain a broader and more complete understanding of architectural design issues in DevOps.

\begin{figure*}[!h]
    \centering
    \includegraphics[scale=0.6]{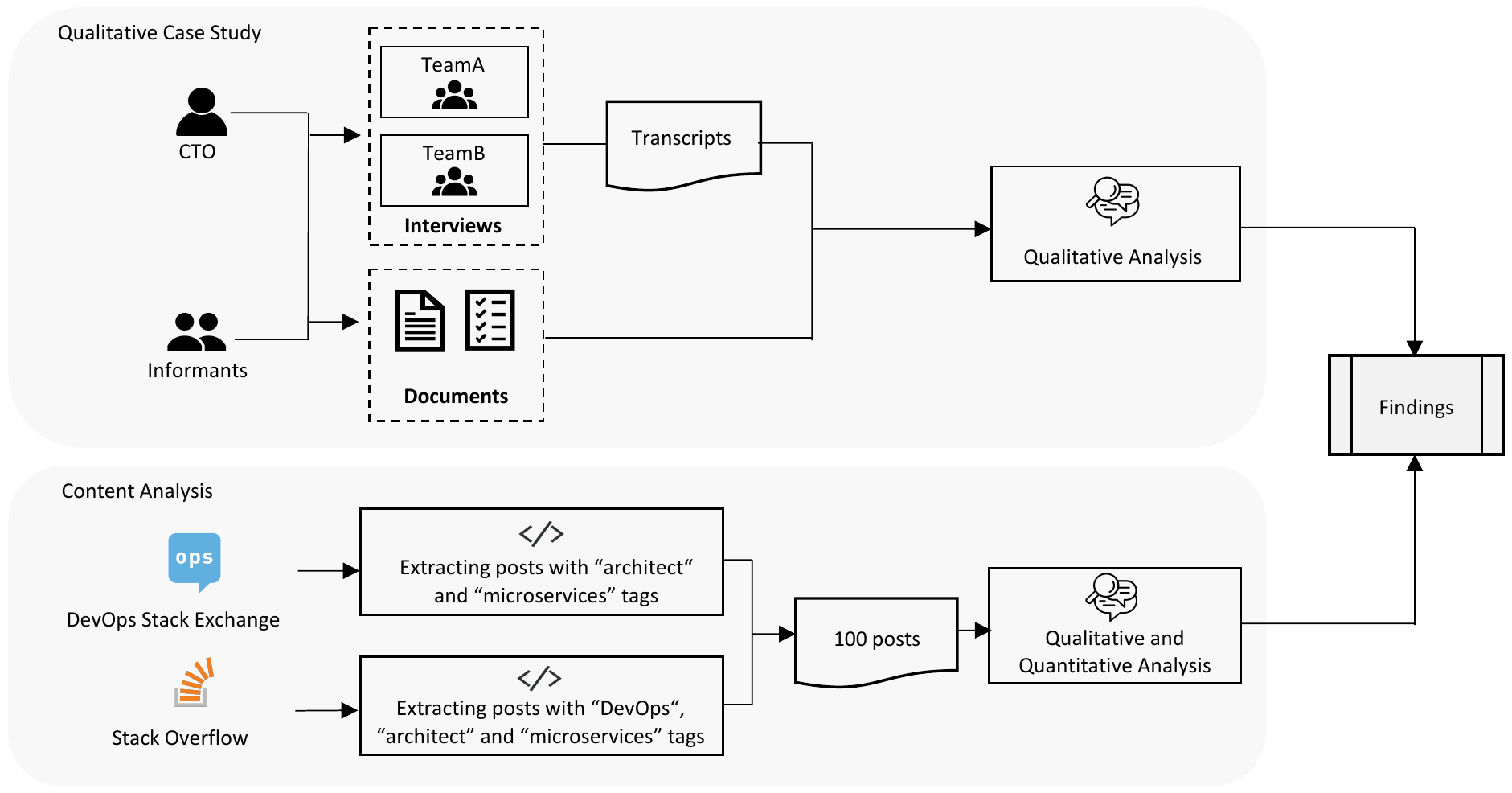}
    \caption{An overview of our mixed-methods study consisting of a qualitative case study and a content analysis of Stack Overflow and DevOps Stack Exchange posts}
    \label{fig:overview}
\end{figure*}

To answer these research questions, we conducted a mixed-methods study consisting of a qualitative case study (\textbf{RQ1}) complemented with a content analysis of Stack Overflow and DevOps Stack Exchange posts (\textbf{RQ2}). As shown in Figure \ref{fig:overview}, the qualitative case study \cite{shahin2020} investigated two teams in a company (henceforth the case company) to understand the specific and contextual architectural design issues faced by them during DevOps transformation and the architectural decisions made to address each design issue. This case study was a holistic, two-case study, as we studied two teams from the same company \cite{yin2017case,prechelt2016quality}. We then collected and analyzed Stack Overflow and DevOps Stack Exchange posts concerning DevOps and architecture to understand architectural design issues in DevOps from a more general perspective. Informed by the established guidelines for conducting empirical studies in software engineering \cite{yin2017case,runeson2009guidelines, ralph2020empirical}, a research protocol was developed in advance and was strictly followed when performing each of the studies.

\subsection{A Qualitative Case Study with Two Teams in a Company (RQ1)}
\subsubsection{Context}\label{seccontext}
\textbf{The Case Company.} The case company is a research and development organization, which develops and delivers robust Big Data solutions and technologies (e.g., tools). By providing such Big Data capability, the customers and end-users of the case company are enabled to make critical decisions faster and more accurately. The case company is made up of several teams working on various research and development programs. Each team includes a variety of roles, such as software engineers, software architects, and data scientists. In this study, we studied two teams: \textbf{TeamA} and \textbf{TeamB}.

\textbf{TeamA.} TeamA develops a social media monitoring platform that collects the available online multimedia data (text, image, and video) and tries to make them digestible to security analysts. This can enable the analysts to extract and identify intelligence and unforeseen insights quickly. Facebook and Twitter are the main data sources for this platform. This project is to descriptively summarize social media content by applying image processing and Natural Language Processing (NLP) approaches. TeamA consists of 8 members, including software engineers, developers, and software architects in a cross-functional team. The team started with four members for about 18 months, but more people were added by growing the project. The platform is a greenfield project. TeamA started with microservices architecture style to have multiple independent deployment units at production, but they changed the architecture of its platform to deploy one monolithic deployment unit at production. The main reason behind this change was the difficulties they experienced during the deployment of a microservices system.

\textbf{TeamB.} TeamB is another team in the case company that works on a greenfield platform. The platform aims at identifying and tracking the potential social event trends. The platform ingests a large amount of publicly available data from a diverse range of social media websites (e.g., Facebook). The goal is to automatically and accurately predict and track society level events such as protests, celebrations, and disease outbreaks. The data science team will then use the predictions. TeamB consists of two teams: one engineering team and one data science team. The work style is that the data science team is the customer for the engineering team, and the data science team has its own customers (e.g., security analysts). The engineering team is composed of 5 members, including system architect and software engineers. The team had recently re-architected its platform from a monolith to a new architecture style (it is called micro architecture), to more rapidly introduce new data sources into the platform. 

\subsubsection{Data Collection}\label{secdatacollection}
The data collection process initially started with an informal meeting. The first and third authors, the CTO of the case company, and two key informants at the case company participated in the informal meeting. This meeting enabled us (the research team) to get a basic understanding of the case company’s structure and domain. It also helped us find the teams adopting DevOps in the case company to be used as a reference point for further steps of our case study. Furthermore, the team members who were suitable for interviews (e.g., those who had a broad view of the software development process, such as software architects and senior software engineers) were identified during the meeting. Finally, the meeting helped us understand what documents and artifacts in the case company should be investigated.

\begin{table*}[!h]
    \centering
    \caption{Projects, teams and interviewees}\label{projectteams}
    \includegraphics[scale=0.6]{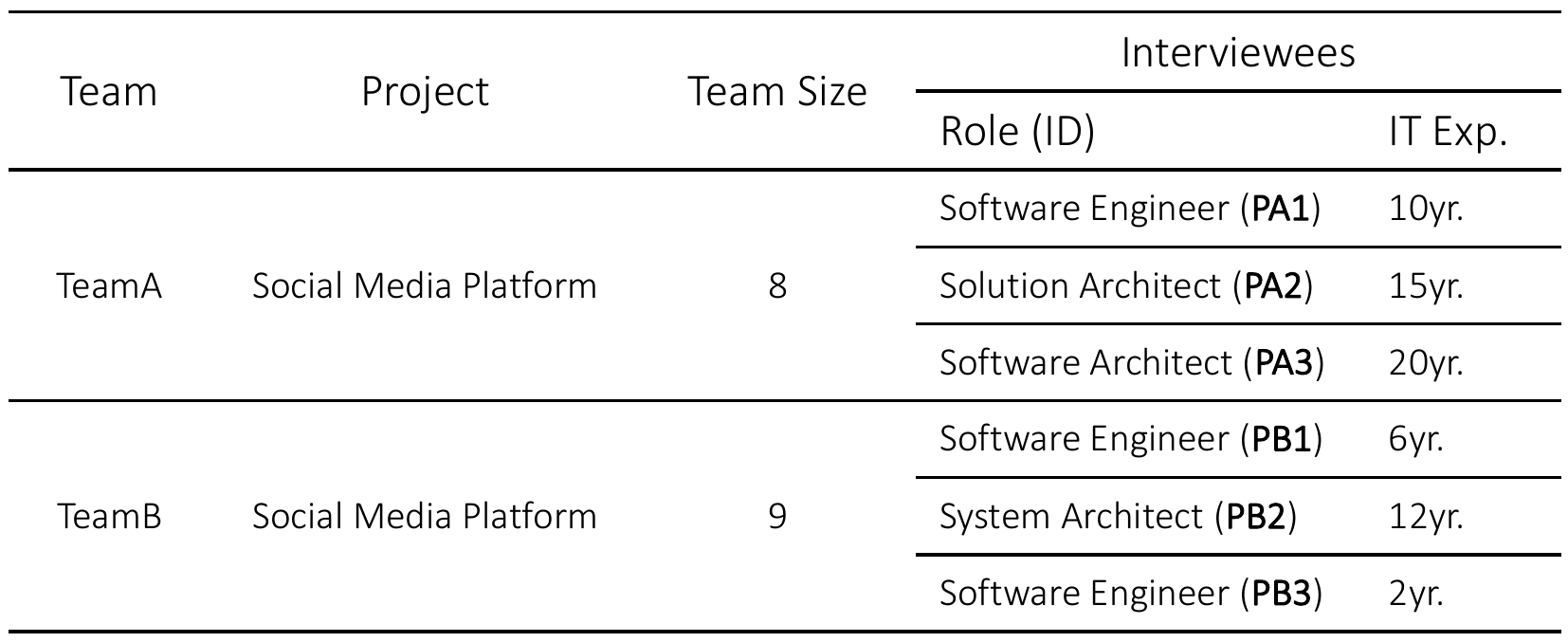}
    
\end{table*}

Face-to-face, semi-structured interviews were the primary tool of data collection. The first author conducted six interviews in total, three interviews with each team. From TeamA, one software engineer, one solution architect, and one software architect participated in the interviews (See Table \ref{projectteams}). Two (senior) software engineers and one system architect from TeamB were also interviewed (See Table \ref{projectteams}). Each interview had 30 open-ended questions, but several follow-up questions were asked based on the participants’ responses. The complete list of the interview questions is available at \cite{onlinedataset}. The initial questions in the interviews were demographic (e.g., participants’ experiences in their current role). The next questions asked about the team’s organization and the characteristics of the project (e.g., domain, deployment frequency, tools, and technologies used for deployment pipeline). For example, it was asked, \textit{\say{Which steps are still manual in your CD pipeline? Why?}} Later, the questions primarily focused on the challenges faced by each team and the practices, decisions, and tools used at the architecture level for adopting DevOps (e.g., \textit{ \say{What architectural characteristics correlate with DevOps?}}). The last part of the interviews investigated the architectural decision-making process in DevOps (e.g.,\textit{ \say{After adopting DevOps in your organization, have you become more independent to make your own (design) decision? If so, how?}}).

However, following semi-structured interviews, the participants were allowed to openly discuss any significant DevOps related experiences and insights they had during their respective projects, not limited to the architecture \cite{hove2005experiences}. It is important to mention that we shared the interview guide with the participants before conducting the interviews. This helped them to be prepared for answering the questions and engaging in discussions \cite{hove2005experiences}. The interviews lasted from 40 minutes to one hour (average: 46.28 minutes, standard deviation: 6.81) and were conducted at the interviewees’ workplaces. All six interviews were audio-recorded with the participants’ permission and then transcribed, resulting in approximately 40 pages of transcripts.

Besides the interview data, the first and third authors had access to more than 120 pages of software development documents provided by the case company and publicly available organizational data (e.g., the case company’s newsletters). Specifically, for each studied team, the first and third authors had access to its project plan document, project vision document, architecture document, and internal team discussions forum. These documents were leveraged to gain a more in-depth insight into important (architectural) design issues, decisions made to meet the design issues in DevOps, and to verify experiences and discussions shared by the interviewees. All of this data was stored on the case company’s wiki and was imported to NVivo \footnote{\url{http://qsrinternational.com}} tool for analysis. 


\subsubsection{Data Analysis}\label{secdataanalysis}

The first author used open coding and constant comparison techniques from grounded theory \cite{glaser1968discovery,hoda2011self} to analyze the interview data and the data collected from the documents. He used NVivo to support qualitative coding and analysis. The first author first created two top-level nodes in NVivo according to our data sources: (1) interview data and (2) document data. Subsequently, open coding was performed with several iterations in parallel with data collection to thoroughly analyze the data gathered from each data source. Key points in the data sources were identified, and a code was assigned to each key point. Listing 1 depicts an example of applying open coding on a portion of an interview transcript. Then, the constant comparison was performed to compare the codes identified in the interviews against each other, as well as to compare them with the codes or the excerpts taken from the documents \cite{hoda2011self}. Next, these emergent codes were iteratively grouped to generate concepts \cite{glaser1968discovery}. Then, the concepts created in the previous step were analyzed to develop categories, which became the architectural design issues presented in Section \ref{sectionStudy1}. As data analysis progressed, the first author constructed relationships among the categories. Then, the identified codes, concepts, and categories were shared with the third author for review and seeking his feedback. The first and third authors held several meetings to discuss the accuracy of the identified codes, concepts, and categories and resolve any disagreements and inconsistencies. The final version of the coded data was agreed upon by the first and third authors.


\begin{table}
\centering
     \textbf{Listing 1} Constructing codes from the interview transcripts

\centering
\begin{tcolorbox}[colback=gray!2!white,colframe=white!80!black]

\textbf{Raw data}: \textit{ \say{We’re trying to make sure everything [to be] more substitutable, which allows to do mocking if we need it. We’re trying to keep everything independent as you can just test that set of function; that succeeded in unit tests}.}
    \\
    \textbf{Key point}: \textit{ \say{Independent stuff as can be mocked and can be independently tested}
    Code: Independent units for test}
    
\end{tcolorbox}
\end{table}

\subsection{A Content Analysis of Stack Overflow and DevOps Stack Exchange Posts (RQ2)}
 
\subsubsection{Building Dataset}
To build a dataset of posts concerning software architecture and DevOps in Stack Overflow and DevOps Stack Exchange, we wrote a query and ran it with the interface provided by Stack Exchange Data Dump \footnote{\url{https://data.stackexchange.com/stackoverflow/query/new}}. The query was designed to retrieve posts with \say{*DevOps*} and \say{*architect*} tags in Stack Overflow and posts with \say{*architect*} tag in DevOps Stack Exchange. The microservices architecture style is considered a promising architecture style to enable DevOps \cite{balalaie2016microservices, zdun2019emerging}. Hence, we also retrieved posts with \say{*DevOps*} and \say{*microservice*} tags in Stack Overflow and posts with \say{*microservice*} tag in DevOps Stack Exchange. For each post, we collected its question title, question body, question's comments, question's answers, and answer's comments. This process led to the collection of 132 posts (65 from Stack Overflow and 67 from DevOps Stack Exchange). We stored them in JSON files and then converted them into HTML files to facilitate our analysis.

\subsubsection{Data Analysis}

The second author applied open coding and constant comparison on the question title, the question body, the question's comments, the question's answers, and the answer's comments of the 132 posts to find architectural design issues concerning DevOps. Note that 32 posts were removed as they were not about architecture and DevOps. Once the second author finished the first round of the data analysis, the first author reviewed all identified codes, concepts, and categories. Then, the first and second authors held several Zoom meetings to discuss the coded data. The second author revised the identified codes, concepts, and categories based on the comments and feedback given by the first author. It should be noted that each post in Stack Overflow or DevOps Stack Exchange may contain several architectural design issues. Our analysis resulted in 11 groups of architectural design issues in DevOps.

Moreover, we examined the popularity and difficulty level of the questions in each group of architectural design issues. To this end, the second author gathered the metadata related to each question. The popularity of each question is calculated based on four metadata: ViewCount (the number of views of a question)\cite{yang2016security, rosen2016mobile, Moses2020Analysis}, FavouriteCount (the number of favorites a question)\cite{yang2016security, Moses2020Analysis}, CommentCount (the number of comments on a question)\cite{yang2016security}, and Score (the score of a question) (See Table \ref{popular}). In line with \cite{yang2016security, Moses2020Analysis}, the average number of views of the questions in a given architectural design issue group can be considered the primary factor for determining the popularity of the architectural design issue group \cite{rosen2016mobile}.


We used three factors to calculate the difficulty of the questions in an architectural design issue group (See Table \ref{difficult}). The first factor calculates the percentage of the questions with no accepted answer in each architectural design issue group \cite{rosen2016mobile, Moses2020Analysis}.
The second factor calculates how long, on average, it takes that the questions in an architectural design issue group receive an accepted answer\cite{yang2016security, rosen2016mobile, Moses2020Analysis}.
In Stack Exchange Data Dump, \say{CreationDate} records the creation date of the questions, and the combination of \say{CreationDate} and \say{AcceptedAnswerId} records the date of receiving an accepted answer. The third factor measures the average number of answers to the questions in an architectural design issue group over the average number of views of the questions\cite{yang2016security}. We refer to this factor as \say{PD} in Table \ref{difficult}. The number of answers to each question is extracted from “AnswerCount”. The number of views of a question in an architectural design issue is recorded by “ViewCount” in Stack Exchange Data Dump. The lower value of the third factor, the more difficult the architectural design issue is.

\section{Findings}\label{secfindings}
In Section \ref{sectionStudy1}, we report the results of the qualitative case study with two teams in a company. Then we discuss the analysis of 100 posts concerning DevOps and architecture from Stack Overflow and DevOps Stack Exchange in Section \ref{sectionSecondStudy}. It is worth mentioning that software architectures in the context of DevOps not only deal with \say{context and requirement} and \say{structure} aspects of software but also should concern about the \say{realization} aspect \cite{hohpe2016software}. The \say{realization} aspect mostly deals with the operational aspects of software \cite{hohpe2016software,woods2016operational}. Moreover, Bass\cite{bass2017software} argues that adopting DevOps necessitates organizational, cultural, and technical changes, in all of which software architect plays a key role. This expands the role and responsibility of architects, as they also need to deal with infrastructure architecture, test architecture, team organization, and automation \cite{shahin2019empirical,hohpe2016software,woods2016operational}. Hence, the architectural design issues in the DevOps context go beyond the application design, and target application architecture, DevOps pipeline architecture, infrastructure architecture, and team organization.

\begin{table*}[!h]
    \centering
        \caption{Eight architectural design issues in DevOps faced by two teams in the case company, their corresponding architectural decisions, and their impacts}

    \includegraphics[scale=0.42]{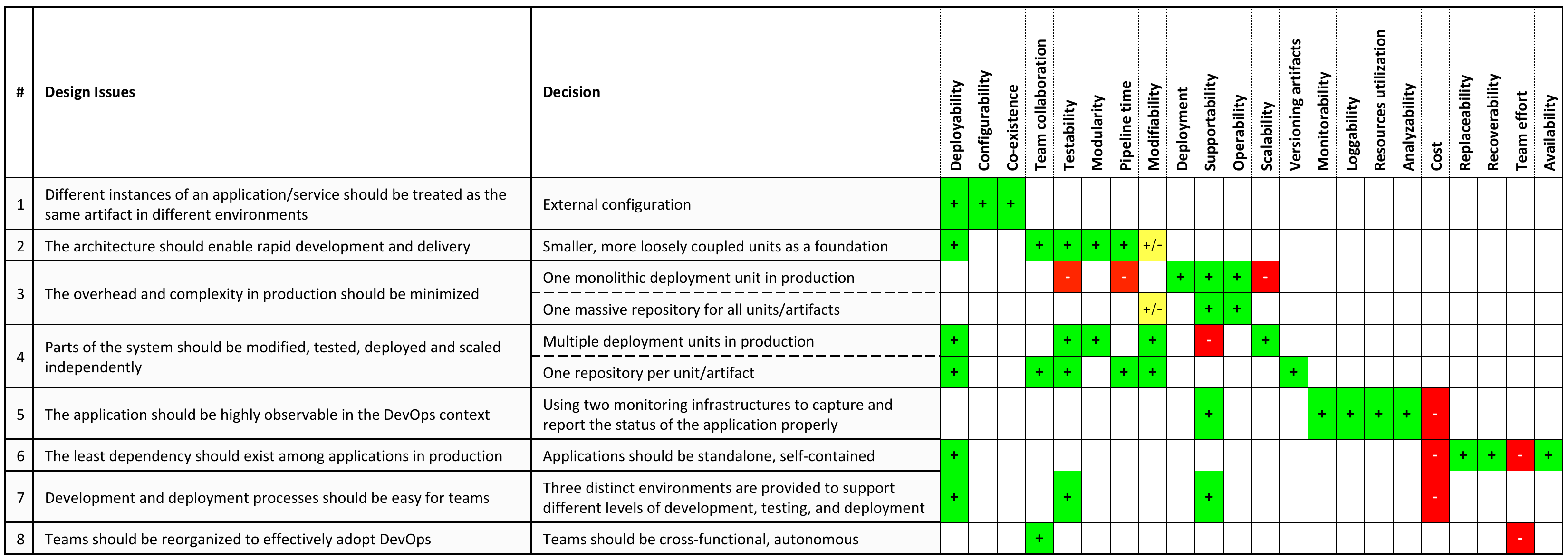}
    \label{tab:implication}
\end{table*}

\subsection{Architectural Design Issues in a Company during DevOps Transformation (RQ1)}\label{sectionStudy1}
\begin{figure*}[!h]
    \centering
    \includegraphics[scale=0.73]{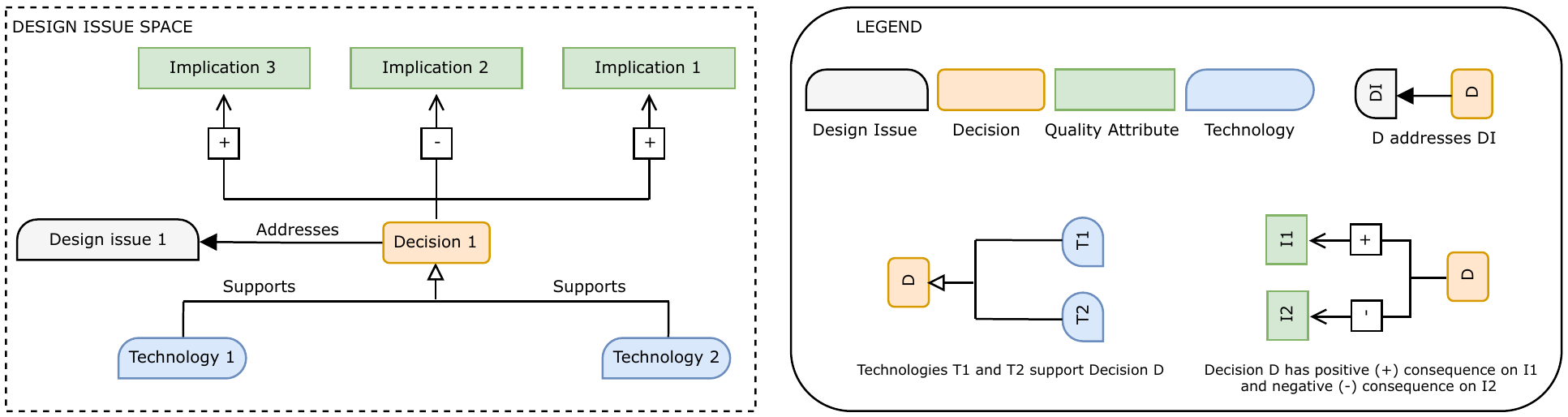}
    \caption{Decision Template Notation}
    \label{fig:notation}
\end{figure*}

\begin{figure}[h]
    \centering
    \includegraphics[scale=0.7]{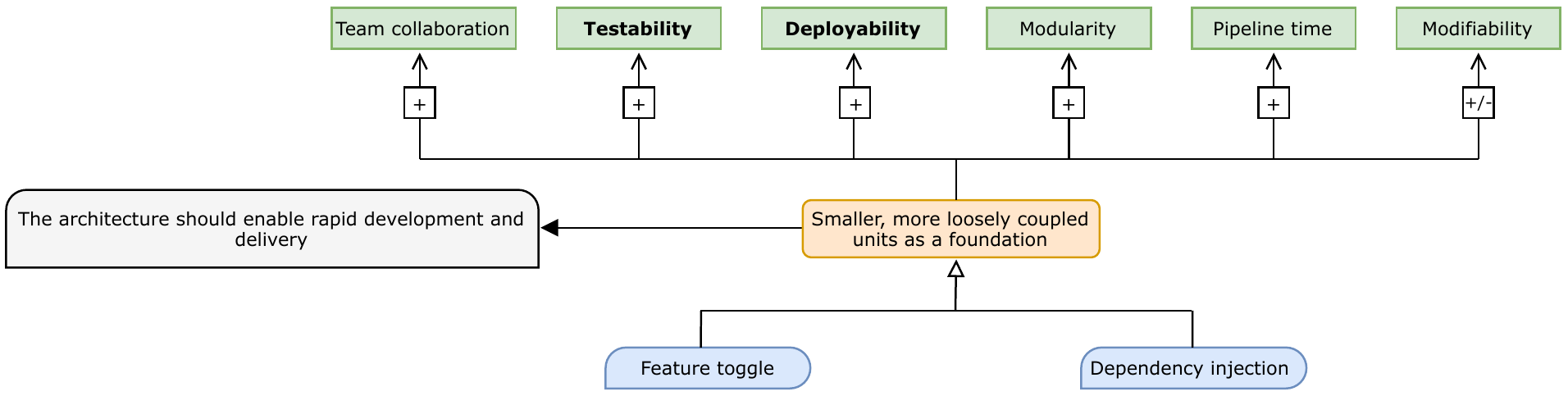}
\caption{Decision template for \underline{Design Issue 2}}\label{fig:Decision2}
\end{figure}

\begin{figure}[h]
    \centering
    \includegraphics[scale=0.7]{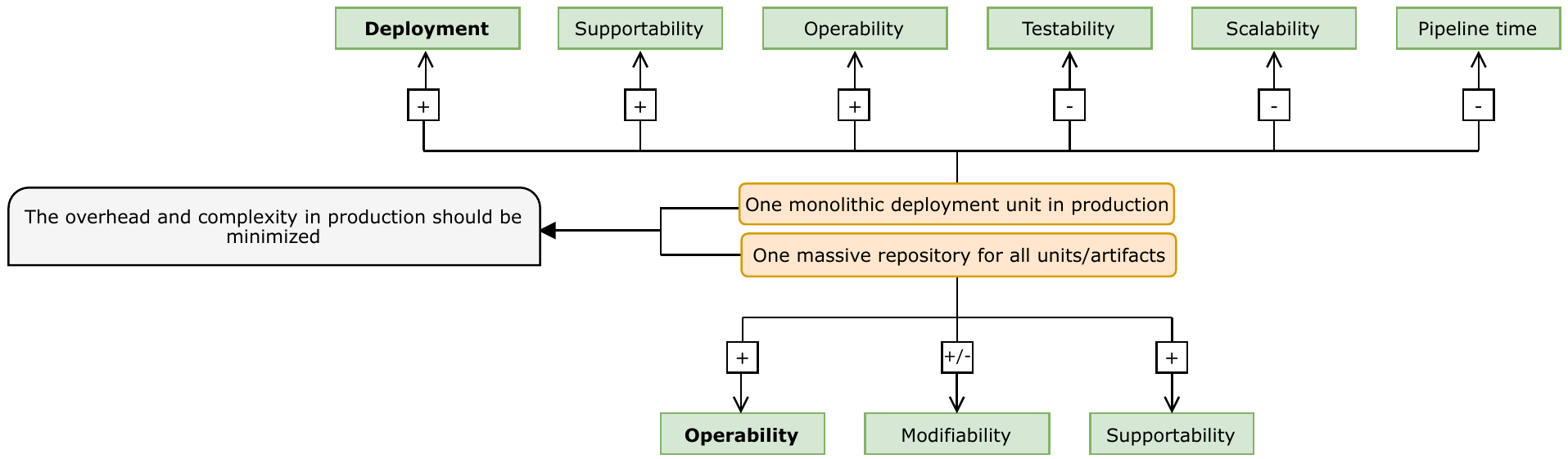}
\caption{Decision template for \underline{Design Issue 3}}\label{fig:Decision3}
\end{figure}

\begin{figure}[h]
    \centering
    \includegraphics[scale=0.65]{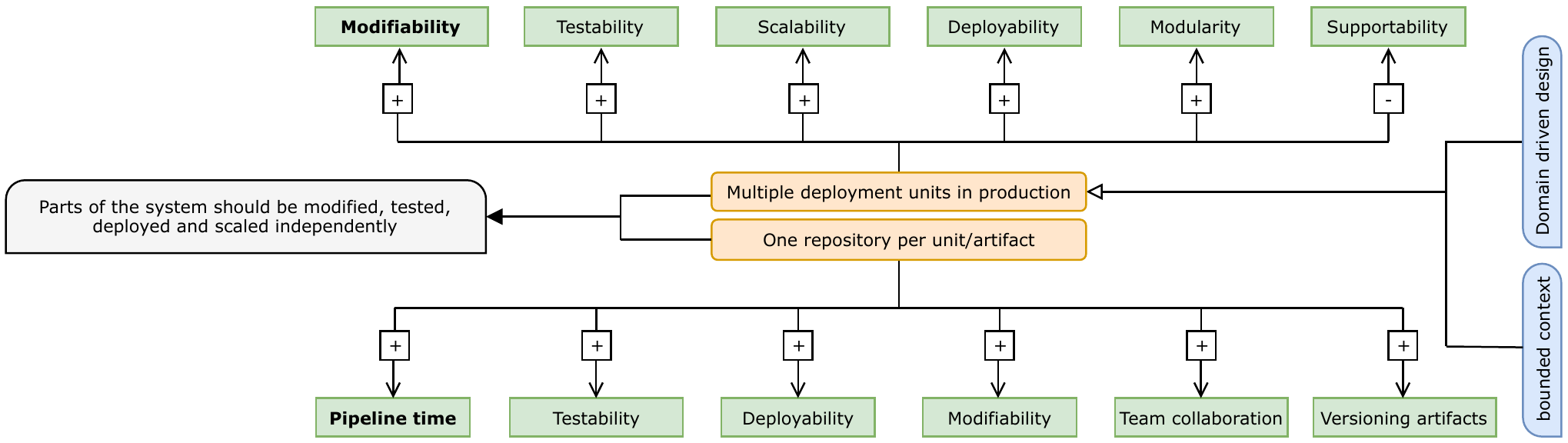}
\caption{Decision template for \underline{Design Issue 4}}\label{fig:Decision4}
\end{figure}

\begin{figure}[h]
    \centering
    \includegraphics[scale=0.7]{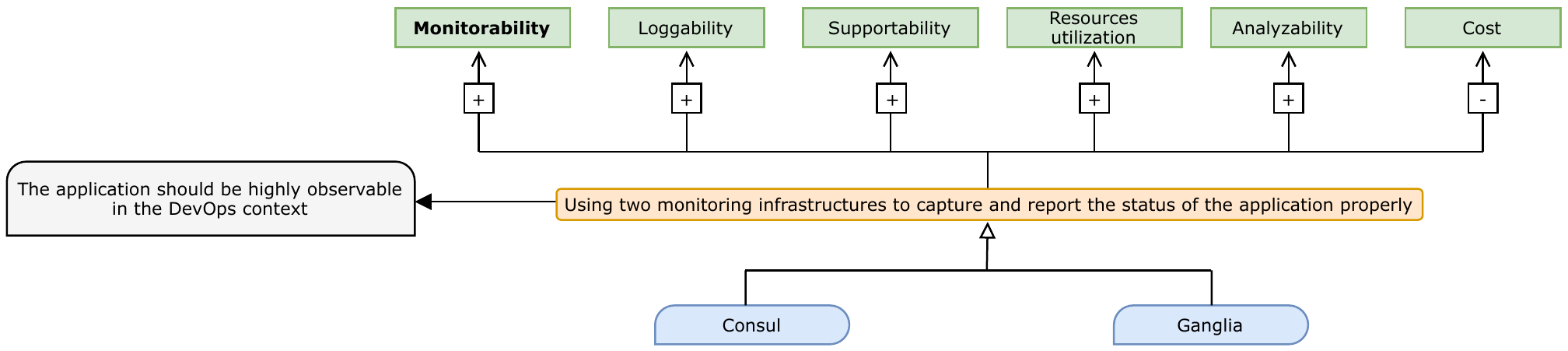}
\caption{Decision template for \underline{Design Issue 5}}\label{fig:Decision5}
\end{figure}

\begin{figure}[h]
    \centering
    \includegraphics[scale=0.7]{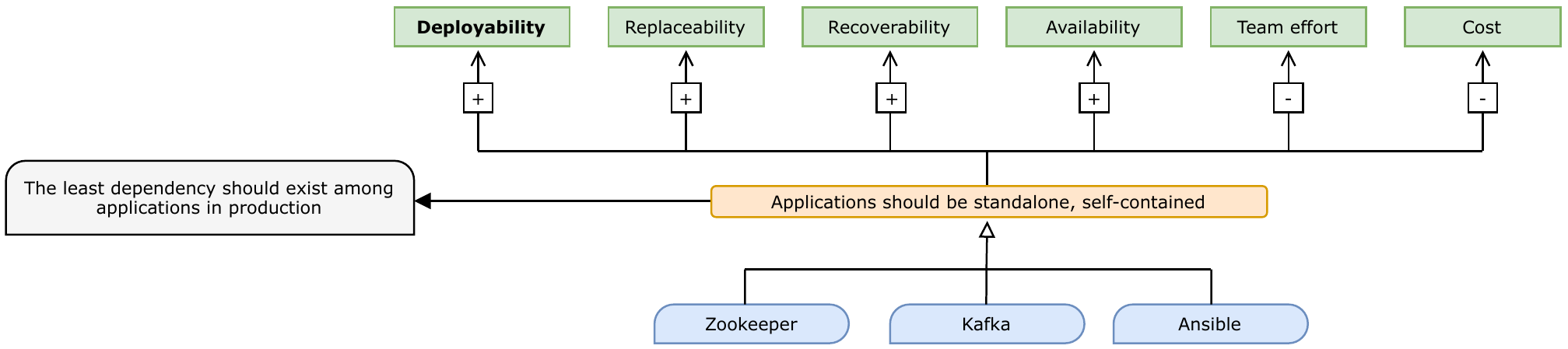}
\caption{Decision template for \underline{Design Issue 6}}\label{fig:Decision6}
\end{figure}

\begin{figure}[h]
    \centering
    \includegraphics[scale=0.75]{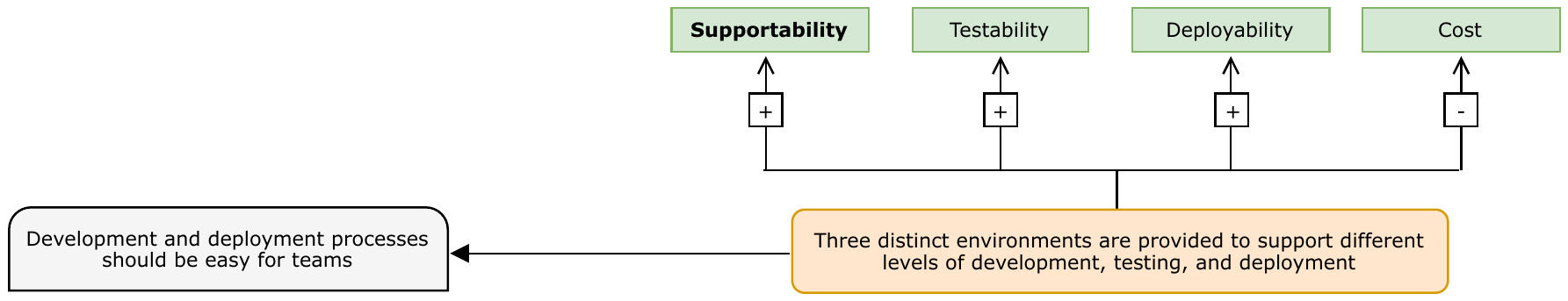}
\caption{Decision template for \underline{Design Issue 7}}\label{fig:Decision7}
\end{figure}

\begin{figure}[h]
    \centering
    \includegraphics[scale=0.75]{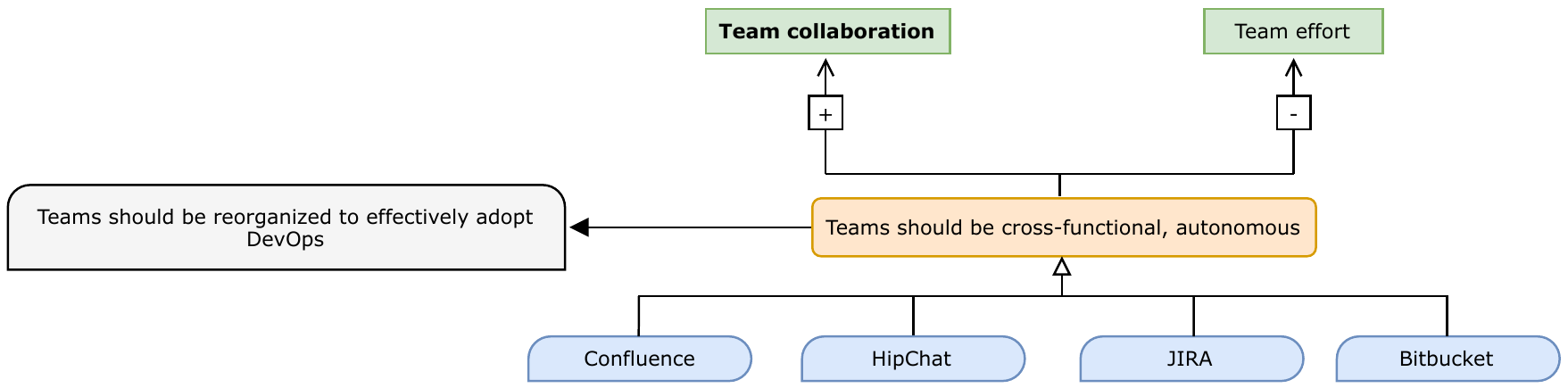}
\caption{Decision template for \underline{Design Issue 8}}\label{fig:Decision8}
\end{figure}

We identified eight contextual and specific architectural design issues and their corresponding architectural decisions of DevOps transformation in the case company. Table \ref{tab:implication} provides an overview of these architectural design issues. Inspired by the decision templates introduced by \cite{haselbock2017decision, lewis2016decision}, we further visualize the identified design issues using a decision template, including design issue, decision, implication, and technology notations. As shown in Figure \ref{fig:notation}, a design issue describes the problem that a decision addresses. A decision may have several implications on different aspects (e.g., quality attributes, team efforts) of software development. Implications can be a positive (represented by \say{+}) or negative (represented by \say{-}) consequence of a decision. \say{+/-} shows a decision might have both positive and negative consequences. Important consequences are \textbf{bold}. Finally, we show any tools, tactics, frameworks, etc., that support (e.g., implement, complement) a decision with the technology notation. Figures \ref{fig:Decision1} to \ref{fig:Decision8} show the decision templates for eight identified architectural design issues. Furthermore, when we refer to a decision in the text, we use \faGavel{} icon, and for technology, we use \faWrench{} icon.

It should be noted that when we refer to data from the interviews with TeamA and TeamB, we use \textbf{PAX} and \textbf{PBX} notations respectively. For instance, \textbf{PA1} refers to interviewee 1 in TeamA (See Table \ref{projectteams}). The excerpts taken from the documents are marked as \textbf{D}.  


\subsubsection{Design Issue 1: Different instances of an application/service should be treated as the same in different environments} \label{sec:DIssue1}
 
\begin{figure}[h]
    \centering
    \includegraphics[scale=0.75]{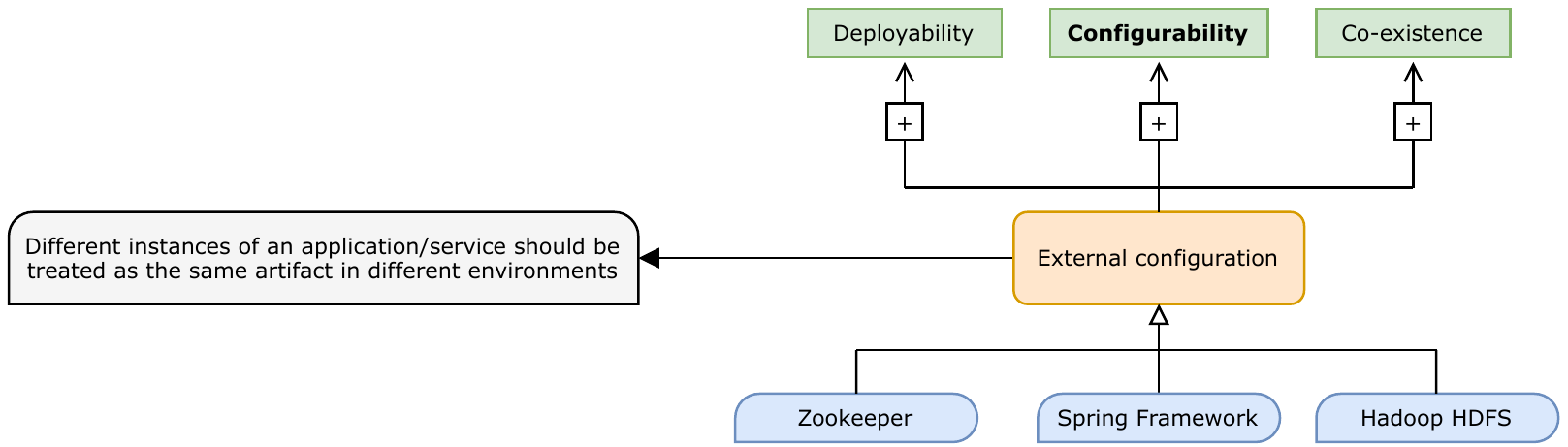}
    \caption{Decision template for \underline{Design Issue 1}}\label{fig:Decision1}
\end{figure}

We have found that the critical design issue in both teams was to ensure that different instances of one artifact in different environments are considered the same artifact. From the interviews and documents, it is clear that the concept of \textit{external configuration} (\faGavel) was adopted to address this design issue (See Figure \ref{fig:Decision1}).
This can be illustrated by this quote: 

\begin{formal}
\faComment{} \textit{ \say{The key architectural thing is [to] making application externally configurable, as you can be aware of the test environment, production environment, etc.}} \textbf{PB2}.
\end{formal}

External configuration aims at making an application externally configurable, in which each application has embedded configuration (i.e., configurations are bundled within an application) for development and test environments and uses the configuration that lives in the production environment. Here configuration refers to storing, tracking, querying, and modifying all artifacts relevant to a project (e.g., application) and the relationships between them in a fully automated fashion \cite{humble2010continuous}. The external configuration also implies \textbf{multiple-level configuration} as each environment has a separate configuration. All this makes deploying applications to different environments trivially easy as there is no complicated setup procedure. By this approach, different instances of one artifact in different environments are considered as the same artifact; once one artifact is deployed to the test environment, the same artifact gets deployed into the production environment. The solution architect from TeamA said: 

\begin{formal}\faComment{} \textit{ \say{We externalize its [application$’$s] configuration as you can provide different [instances] in different environments but as the same artifact}} \textbf{PA2}.
\end{formal}

Apart from the positive impact on deployability, this decision leads to improving configurability. This is because the embedded configurations inside applications can be easily overwritten at target environments, and there is no need to reconfigure the whole infrastructure. Those configurations that might rapidly change are read from \textit{Zookeeper} (\faWrench{}) \footnote{\url{https://zookeeper.apache.org}}, but large and static ones are read from \textit{HDFS} (\faWrench{}) \footnote{\url{https://hadoop.apache.org/docs/r1.2.1/hdfs\_design.html}} (Hadoop Distributed File System). \textbf{PB2} explained the benefits of \say{external configuration} decision vividly: 

\begin{formal}\faComment{} \textit{ \say{We also use Spring Boot. It is always looking for module name first, then for the file directly next to JAR, and then for the embedded JAR. We have multiple levels of configuration in our JAR file. So, inside our JAR, we have the same default, and they always target the local environment. If you accidentally run a JAR file and you might do something crazy like delete data, it lonely targets your local environment.}}
\end{formal}


\subsubsection{Design Issue 2: The architecture should enable rapid development and delivery}

Our participants confirm that a DevOps-driven architecture needs to be loosely coupled to ensure team members working on would have a minimum dependency. By this, developers can get their works (e.g., development) done with high isolation \cite{forsgren20172017}. Hence, both teams (A\&B) consider \textit{smaller, more loosely coupled units as a foundation} (\faGavel) for architecting in DevOps. For example, we have:

\begin{formal}\faComment
\textit{ \say{I guess [we are] trying to be mindful of reducing coupling between different parts of the application as we can separate those things}} \textbf{PB1}.
\end{formal}

\begin{formal}\faComment
\textit{ \say{[In this new architecture,] we want to be able to address each of these stories, without being tightly coupled to another}} \textbf{D}.
\end{formal}

The interviewees reported that this decision is a substantial benefit to testability and deployability. If decoupled architecture is fulfilled, it appears that team members can independently and easily test units (e.g., better test coverage) and drastically decrease the cycle time of test and deployment processes. \textbf{PA1} affirmed that they were successful in implementing this principle as everything for \say{unit tests} is independent and substitutable, which allows them to do mocking if needed. In both projects, the interviewees explained that the decoupled architecture was achieved by extensive use of \textit{dependency injection} (\faWrench{}) \footnote{\url{https://martinfowler.com/articles/injection.html}}, \textit{feature toggle} (\faWrench{}) \footnote{\url{https://martinfowler.com/articles/feature-toggles.html}} , and building units that are backward and forward compatible. A participant from TeamA it like this: 

\begin{formal}\faComment
\textit{ \say{In terms of architecture, [we] build things that they are decoupled through interfaces, using things like dependency injection, using feature flags; these are [the topics related to] the architecture that we use to support deployability}} \textbf{PA2}.
\end{formal}

This decision enabled the teams to make sure everything is nicely separable, reconfigurable, and extensible. The participants shared that all self-contained applications and components were intentionally designed small to be tested in isolation. As an example:

\begin{formal}\faComment
\textit{ \say{Number one thing in DevOps is to have software to be well tested, which you need to separate concerns into separate components that you can test individual piece of functionality [without] influencing other components}} \textbf{PB1}.
\end{formal}

\textbf{PB2} discussed the benefit of \say{smaller, more loosely coupled units} from a different perspective. According to him, breaking down a large component (called Enricher) into five smaller and independent units enabled them to increase test coverage of each to 90\%. \textbf{PA3}, on the other hand, pointed out that this was also helpful for having a more efficient DevOps pipeline as large components increase DevOps pipeline’s time and are hurdles for having quick and direct feedback from test results. 


\subsubsection{Design Issue 3: The overhead and complexity in production should be minimized}
Whilst TeamA and TeamB had many common architectural decisions (e.g., external configuration) in the context of DevOps, they deal with the deployment process differently. This significantly impacted on their architectural decisions. TeamA had started with the microservices architecture style. Still, once the team felt difficulties in the deployment process, they switched to \textit{a monolith to minimize the number of deployment units in operations} (\faGavel). This significantly helped them to address the deployment challenges encountered earlier with the microservices architecture style. \textbf{PA2} and \textbf{PA3} explained it in this words:

\begin{formal}\faComment
\textit{ \say{We actually started with microservices. The reason was that we wanted to scale out some analysis components across machines. The requirements for the application changed, and this led to [move to] this monolith.}}
\end{formal}

\begin{formal}\faComment
\textit{ \say{We deploy it [the application] as one JAR file.}}
\end{formal}

We found the following reasons for building a monolithic-aware deployment: (i) It is much easier to deploy one JAR file instead of deploying multiple deployment units. This mainly because the deployment units are always required to be locked in the end, as some changes are made to all of them at the same time. (ii) Having multiple deployment units can increase overhead in operations time. When \textbf{PA1} was asked why their system (i.e., platform) uses one monolithic deployment, he replied:

\begin{formal}\faComment
\textit{
\say{There was a lot of overhead in trying to make sure that this version is backward compatible with this version of that one.}}
\end{formal}

(iii) It can be difficult to manage changes in the development side, where there are multiple, independent deployment units. In this scenario, TeamA found that it is easier to bundle all units together into one monolithic release. \textbf{PA1} added:

\begin{formal}\faComment
\textit{ \say{We are doing monolith. I think that$’$s been slowly changing idea; we were originally going to have several suites of [deployment units] that worked together, but I think to cut down on development effort, we are just going to put all in one tool.}}
\end{formal}

Furthermore, TeamA decided to use \textit{one repository for all modules and libraries} (\faGavel). This decision was heavily influenced by its monolithic design. The project of TeamA used to have multiple repositories for some of its components and libraries. This caused challenges in the deployment process, such as it made it harder to synchronize artifact’ changes, and it presented the integration problem. \textbf{PA3} explained: 

\begin{formal}\faComment
\textit{ \say{If you make a change to one component, it is still built and passed the tests, but when they would need to be together, it wouldn$’$t work. So, we merged all the components in the same repository.}}
\end{formal}


\subsubsection{Design Issue 4: Parts of the system should be modified, tested, deployed and scaled independently}

The opposite scenario happened to TeamB as they had recently re-architected its monolithic system to have multiple deployment units. 

\begin{formal}\faComment
\textit{ \say{We cannot have a thing like TeamA [as] they have one monolithic JAR, and everything is just in there and just deploy to one post. We are very much more microservices; we like to be far away from that [monolithic deployment] as much as possible}} \textbf{PB2}.
\end{formal}

This new architecture, which is called micro architecture by TeamB, is a combination of microservices and monolith approaches. Among others (e.g., parts of TeamB’s platform should scale and be tested independently of others), we observed that evolvability (modifiability) was the main business driver for this transition.

\begin{formal}\faComment
\textit{ \say{We used to have one big monolithic application, and changing anything required to the redeployment of the whole application, which interrupted all processes. [It is mainly] because we couldn$’$t change anything really without taking down others}} \textbf{PB1}.
\end{formal}

TeamB used \textit{bounded contexts} (\faWrench{}) and \textit{domain-driven design} (\faWrench{}) \cite{evans2004domain} as a way to split large domains, resulting in \textit{smaller, independent, and autonomous deployment units (e.g., microservices and self-contained applications)} (\faGavel). TeamB’s system includes more than forty libraries, microservices, and autonomous applications that are built and deployed automatically and independently. They are very small, and the intention was to keep them small as each of them should be \textbf{single bounded} and should do \textbf{only} one specific task. This enabled them to minimize external dependencies. \textbf{PB2} reported: 

\begin{formal}\faComment
\textit{ \say{We have seven Ingestion apps running; every single app has very specific things to do, like Twitter Ingestor [app] only does Tweets.}}
\end{formal}

Driven by the team’s intention to build, test, and deploy the artifacts in isolation, TeamB decided to build and maintain \textit{one repository per each artifact} (\faGavel). They were frustrated by keeping all artifacts into one repository because with the monolithic repository, changing one thing required running all tests and re-deploy all things. \textbf{PB3} referred to this decision as a key architectural decision that was deliberately made to simplify DevOps transformation. This helped them manage the versioning of different libraries, modules, and self-contained applications in the deployment process. Interviewee \textbf{PB1} had a similar opinion to \textbf{PB3} and said: 

\begin{formal}\faComment
\textit{ \say{The number one thing is that reusable libraries or components should potentially be in own artifact. So, you can build, test, and deploy that artifact in isolation. With the monolith [repository], you change one thing to test that, the monolith is going to run all tests for everything for being (re) deployed.}}
\end{formal}

We discovered that the concern about the time taken by the DevOps pipeline was another reason for this decision. \textbf{PB3} described it as \say{the build cycle in our old architecture was problematic as it hit the development and iteration}. In the previous architecture, only the build cycle took around 10 minutes. By applying the above changes, the cycle time of the pipeline turned around as currently all the build time, Ansiblizing the deployment process, and release to Nexus \footnote{https://www.sonatype.com/nexus-repository-sonatype} take approximately 10 minutes.


\subsubsection{Design Issue 5: The application should be highly observable in the DevOps context}

Monitoring and logging were identified and discussed by all interviewees as a primary area, which needs more attention in the DevOps context. In both projects, they build logs and metrics data into software applications as monitoring tools could leverage these data. Consequently, the applications in the DevOps context need to capture and report data about their operations properly \cite{bosch2015speed}. The system architect in TeamA (PA3) believed that this is the most important implication driven by DevOps in terms of architecture. Another interviewee described the relationship between architecture and monitoring as the following: 

\begin{formal}\faComment
\textit{ \say{In order to trust your ecosystem works especially when you are changing things to be continuously deployed; if you don$’$t have strong architecture, then the application$’$ changes are less clear and is more difficult to monitor}} \textbf{PB1}.
\end{formal}

Both TeamA and TeamB use two \textit{monitoring infrastructures} (\faGavel), \textit{Consul} (\faWrench{}) and \textit{Ganglia} (\faWrench{}), which are shared among a couple of projects in the case company. These systems are used for different purposes. \textit{Consul} (\faWrench{}) \footnote{https://www.consul.io/} is mostly used as a high service to check the state of an application. This monitoring system aggregates and orchestrates metrics for cluster state, with which both the teams can identify components’ changes among critical, warring, or good states. According to an interviewee, this enables them to prioritize stability issues in clusters, and from that, they are able to implement new fixes and deploy those fixes and capture the relevant information. In contrast, \textit{Ganglia} (\faWrench{}) \footnote{http://ganglia.sourceforge.net/} is used for aggregating metrics like disc usages, input and output network, interfaces, memory usage, CPU usage. \textbf{PB2} described: 

\begin{formal}\faComment
\textit{ \say{It [Ganglia (\faWrench{})] is kind of aggregating the information. For example, this cluster node is constantly 100\% CPU usage; this cluster almost is full disc usage. We can identify those stability issues as well and find out them early before becomes a real problem}} \textbf{PB3}.
\end{formal}

It appears that whilst both the teams extensively improve their logging and monitoring capability using \textit{Consul} (\faWrench{}) and \textit{Ganglia} (\faWrench{}) to become DevOps, they are not interested in metrics data analytics. Apparently, the log and metrics data are mostly used for diagnostic purposes, and they are not used for an introspective purpose (i.e., understanding how an application/service can be improved based on runtime data). We found that the low number of end users was the reason why TeamA and TeamB are not interested in metric analysis. 

\begin{formal}\faComment
\textit{ \say{We don$’$t do a lot of analysis of the metrics. We do collect a lot of metrics, [for example] we have some centralized logging and centralized monitoring servers and pull all stuff in there, but it is mostly for diagnostic if the application fails and trying to figure out what went wrong}} \textbf{PA1}.
\end{formal}

In both projects, the abundance of monitoring and logging data produced by the applications presents severe challenges, in which scaling up machines and capacities cannot solve the problem anymore. As a result, most of the participants felt that they have to re-think their logging and monitoring approaches and how to reduce the size of logs. One participant complained about this pain point: 

\begin{formal}\faComment
\textit{ \say{We$’$ve got Consul to let us know memory usage in that computer. We$’$ve got a server, which executed by multi-tenant solution, and each of them is running their own application that may spin up in parallel; that consumes memory. So, you get the capacity issues. So maybe the solution was to scale out machine initially, but now I think that is an issue again}} \textbf{PB1}.
\end{formal}


\subsubsection{Design Issue 6: The least dependency should exist among applications in production}
We found frequent references in the interviews to how the applications can get easily and reliably deployed to different environments. Besides applying \say{external configuration}, both the teams decided that \textit{the deployment of an application or service be locked and independent of other applications and services that it depended on} (\faGavel). To this end, they have adopted a \textbf{self-contained} and \textbf{self-hosting} approach \cite{cerny2017disambiguation}, in which all dependencies need to be bundled to an application. An interviewee described the impact of this approach as: 

\begin{formal}\faComment
\textit{ \say{Each of them [applications] like Ingestors are considered as a single standalone, self-contained application, and the pipeline is a self-contained app. They don$’$t really talk with any things else; they don$’$t have any external dependencies like taking one of those down doesn$’$t affect the other ones}} \textbf{PB2}.
\end{formal}

To avoid the interruptions of self-contained applications at runtime, they provide, where possible, the necessary infrastructures (e.g., load balancers) and multiple instances of applications or services behind them. Using Apache \textit{Kafka} (\faWrench{}) \footnote{https://kafka.apache.org/} also enables them to segregate service components and isolate them as a service component can be easily swapped in/with new ones at operations: 

\begin{formal}\faComment
\textit{ \say{By using Kafka Buffer, you take a component out and then message its Buffers to either redeployed artifacts or new artifacts that are connected and consume messaging}} \textbf{PB3}.
\end{formal}

\textit{Ansible} (\faWrench{}) \footnote{https://www.ansible.com/} tool is used as the main automation deployment tool to make sure everything is deployable using infrastructure-as-code. Our interviewees reported two main methods to achieve this. First, it should be ensured that everything, along with its dependencies that are going to be deployed, should be captured in \textit{Ansible} (\faWrench{}). Second, Ansible playbooks should be written well. From TeamB’s architect perspective, writing good Ansible playbooks was the biggest challenge in their DevOps journey.

In TeamB’s project, everything is stateless as well. In addition, data fields are designed optional, which they can either exist or null. This was indeed helpful to the deployment process because there is no need to have the right database with the right data in it to restart the self-contained applications. TeamB extensively uses \textit{Zookeeper} (\faWrench{}) \footnote{https://zookeeper.apache.org/} tool to make code stateless. \textbf{PB2} explained: 

\begin{formal}\faComment
\textit{ \say{There is a clean separation between the apps. So, I can go and update Ingestors; I can go and updates DTO and add more data or meta-data to it without affecting anything downstream; like I just continue operating on the old version of schema.}}
\end{formal}


\subsubsection{Design Issue 7: Development and deployment processes should be easy for teams}
Providing sufficient physical resources (e.g., CPU and memories) to both the teams enabled them to \textit{establish three distinct environments, including development, integration (or test), and production environments} (\faGavel). In both teams, this was indeed a conscious decision to manage different levels of development, testing, and deployment in DevOps transformation \cite{bass2015devops}. 

\begin{table}[h]
\caption{How often the teams deploy to different environments}\label{diffenv}
\centering
\begin{tabular}{l|c|c|c|}
\cline{2-4}
                                                    & \multicolumn{3}{c|}{Deployment Frequency to}                                \\ \hline

\multicolumn{1}{|l|} {}    & Development Env.    & Integration Env.         & Production Env.            \\ \hline
\multicolumn{1}{|c|}{TeamA} & Multiple-time a day & Once per week            & Once per Sprint (two-week) \\ \hline
\multicolumn{1}{|c|}{TeamB} & Multiple-time a day & Multiple-time per Sprint & At least once per Sprint   \\ \hline
\end{tabular}
\end{table}
Table \ref{diffenv} shows the frequency of deployments in these environments is different. They strictly follow three upfront rules in the above-mentioned environments to ease DevOps adoption: (i) in the development environment, \textbf{only} unit tests need to be performed, and the integration environment should include \textbf{all dependencies} to properly run integration tests against all snapshot builds. One participant described the test environment like this:

\begin{formal}\faComment
\textit{ \say{You can bring to your own laptop all the infrastructure requirements of the cluster like Kafka Buffer, Postgre databases, and Zookeeper instances. Then you locally test against that, and then Jenkins automatically tests against the test environment}} \textbf{PB1}.
\end{formal}

(ii) Code reviews should be performed \textbf{before} committing to the development branch. TeamA and TeamB use the Gitflow branching model\footnote{https://bit.ly/2HYjIfB} , in which developers do their work on feature branches, and before merging changes to the development branch, they should raise a pull request to review the changes in \textit{Bitbucket} (\faWrench{}) \footnote{https://bitbucket.org/} \cite{gousios2016work}. Then that pull is merged into the development branch. Except for large tasks that might not be merge-able in few days, multiple pull requests occur a day. The interviewee \textbf{PA2} reported that these rules \say{improve the merge [quality], and [make] development branch builds always ready for snapshot and deploying to the test environment}.

(iii) Critical bugs should be fixed \textbf{only} on the release candidate branch, not on the development branch. This working style, along with automation support, was deemed helpful to faster repair bugs and reduce the risk of deployed changes to production. 

\begin{formal}\faComment
\textit{ \say{If there is an issue, we can quickly fix and redeploy quickly. So, the turnaround is very small because all are automated at the moment}} \textbf{PB3}.
\end{formal}

Applying the above rules leads to the development branch being potentially in the releasable state anytime, but it is not necessarily stable. An interviewee summarized it as:

\begin{formal}\faComment
\textit{ \say{It [main branch] is a stable artifact. The development branch, which is all integrated features, ready to be merged into master anytime, [but] that might not be quite stable}} \textbf{PB3}.
\end{formal}

Although teams deploy to production at the end of a Sprint, actual production deployments are not tied to the Sprint and can be frequent.

\begin{formal}\faComment
\textit{ \say{We are working two weeks Sprint. We always try to our releasable is done in two weeks Sprint but often [we have] more releases, quite often. I think it is releasable in a couple of times a week}} \textbf{PB1}.
\end{formal}

\begin{formal}\faComment
\textit{ \say{Releases are not tied to Sprint tempo they can be more or less frequent}} \textbf{D}.
\end{formal}


\subsubsection{Design Issue 8: Teams should be reorganized to effectively adopt DevOps}
Another key decision made by the studied organization to support DevOps transformation is to set up \textit{small, cross-functional teams} (\faGavel). Both of the studied teams have \textbf{end-to-end ownership} of their products, from design to deployment to support the code in production: 

\begin{formal}\faComment
\textit{ \say{The significant part of the engineering team$’$s time is to maintain DevOps aspect of the app$–$ I mean the complexity of operations as we have to maintain infrastructure as code and understand that, we have to spend time looking at metrics}} \textbf{PB1}. 
\end{formal}

Furthermore, each team member should perform \textbf{testing} and \textbf{operations tasks} (e.g., writing Ansible playbooks). In both the teams, testing is considered as a rotating activity, not as a phase, which should be performed by all developers. It should be noted that some software engineers have been frustrated with doing operations tasks, as they believe this can be a source of distraction for doing real software development tasks (e.g., debugging). As an example:

\begin{formal}\faComment
\textit{ \say{Sometimes you [as developer] spend a lot of time on debugging Ansible and Jenkins, infrastructures things, for example why Kafka Buffer is not coming from this host. It is kind of being dust by DevOps on deployment task}} \textbf{PB2}.
\end{formal}

Both teams are also \textit{\textbf{autonomous} (\faGavel)} in terms of having the freedom to act independently and make their own architectural and technology decisions. For instance, as discussed earlier, they have chosen two different approaches for architecting their respective systems. \textbf{Collaboration culture} is well established in both teams through \textbf{co-design} and \textbf{shared responsibilities}. Regarding the partnership in testing activity, the interviewee \textbf{PA1} added: 

\begin{formal}\faComment
\textit{ \say{We are rotating testing role, so one person is responsible for testing everything in the previous Sprint and then press proof button and say: to get deployed to the production.}}
\end{formal}

Designing and evaluating systems is mainly performed by the architects in each team. However, the architects regularly code and test, and other team members actively and substantially participate in the architecting process. The architect of TeamA put the collaborative design in these words: 

\begin{formal}\faComment
\textit{ \say{Everybody is involved in architecture [design]. We are not going to have only I as a solution architect; everyone in the team can do that. The architecture is done in daily work, and everyone can have his [own] idea}} \textbf{PA2}.
\end{formal}


\begin{figure*}[b]
    \centering
    \includegraphics[scale=0.42]{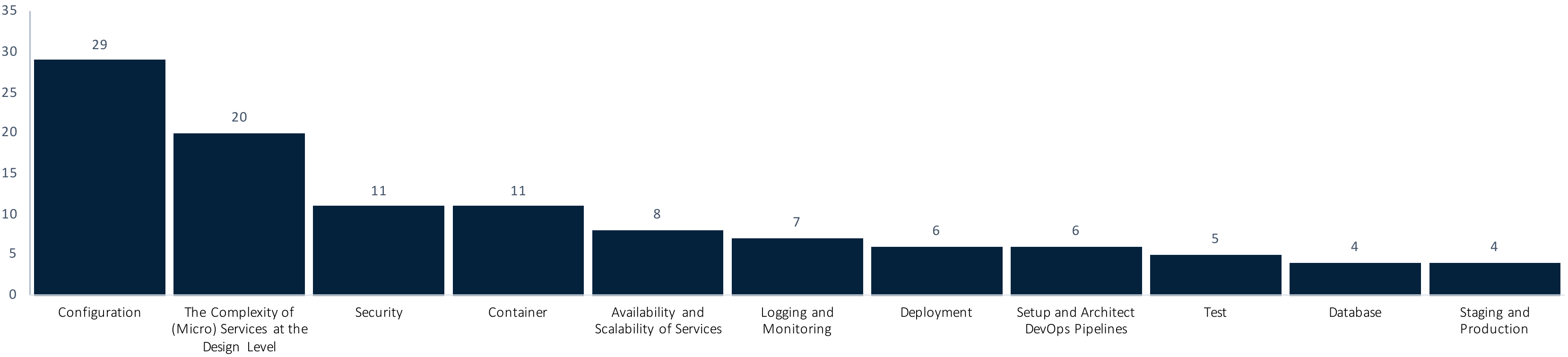}
    \caption{Number of posts in each architectural design issue group}
    \label{fig:number_of_posts}
\end{figure*}

\subsection{Architectural Design Issues in Q$\&$A Websites (RQ2)}\label{sectionSecondStudy}
This section introduces 11 groups of architectural design issues in DevOps raised and discussed by practitioners in Stack Overflow and DevOps Stack Exchange. The architectural design issues come from a qualitative analysis of 100 posts collected from Stack Overflow and DevOps Stack Exchange. Figure \ref{fig:number_of_posts} sorts these 11 architectural design issue groups based on the number of posts in each category. We describe each architectural design issue group with a few representative posts. We also show the level of popularity and difficulty of architectural design issues and the posts in each architectural design issue group. Table \ref{popular} and \ref{difficult} show the popularity and the difficulty of architectural design issues, respectively.

\subsubsection{Group 1: Configuration}
Our analysis shows that configuration is the fourth most popular architectural design issue (835 average number of views and 1.28 average number of comments) and has the most number of questions (29 out of 100 posts).
This category is mainly about configuring a wide range of DevOps tools (e.g., configuring \textit{Ansible} (\faWrench{}) for runit \footnote{runit is a cross-platform Unix init scheme with service supervision.}) and technologies (e.g., configuring Docker files and configuring Pods) and (micro) services in DevOps pipelines. The difficulty level of this design issue group is relatively high, as 58.62\% of the questions in this design issue group do not receive accepted answers. The most viewed question (8000 views) in this category is: 

\begin{formal}\faComment
\textit{ \say{My understanding of Ansible roles is that they are the unit of reusability when implementing processes using \textit{Ansible}. [...] It therefore seems natural to use roles to make all tasks understanding a given data structure in a specific role, but this yields the question of how to select the actual tasks to perform? Is it a good design to use the main.yml file in the task as an entrypoint to select actual tasks according to role parameters?
}} \footnotemark
\end{formal}\footnotetext{\url{https://devops.stackexchange.com/questions/3498}}

Another question seeks recommendations for configuring Jar files using Ansible playbook: 

\begin{formal}\faComment
\textit{ \say{[...] We intend to configure the executable jar files to be supervised by runit, instead of SysVinit or Upstart. Is it reasonable to package the runit config inside the RPM, and let \textit{Ansible} only deploy specific properties into /etc/sysconfig, or does a better way to use \textit{Ansible} for the full runit config exist, so that the service would be decoupled from that host configuration?}} \footnotemark
\end{formal}\footnotetext{\url{https://devops.stackexchange.com/questions/105}}

\begin{table*}[t]
\caption{Popularity of architectural design issue groups}\label{popular}
\centering
\addtolength{\tabcolsep}{-4pt}
\begin{tabular}{lccccc}
\hline
\textbf{Architectural Design Issue}                    & \textbf{\# of Posts} & \textbf{Avg. Views} & \textbf{Avg. Favorites} & \textbf{Avg. Comments} & \textbf{Avg. Scores} \\ \hline
Availability and Scalability of Services               & 8                    & 2280.37             & 2.37                    & 1.5                    & 7.12                 \\
Test                                                   & 5                    & 1266.8              & 2.2                     & 1                      & 4                    \\
Setup and Architect DevOps Pipelines                   & 6                    & 1027                & 1.33                    & 0.33                   & 2                    \\
Configuration                                          & 29                   & 835                 & 0.52                    & 1.28                   & 2.41                 \\
The Complexity of (Micro) Services at the Design Level & 20                   & 496.2               & 0.5                     & 0.8                    & 2.05                 \\
Security                                               & 11                   & 462.09              & 0.45                    & 1.64                   & 3.18                 \\
Deployment                                             & 6                    & 420.5               & 0.17                    & 2.17                   & 2                    \\
Container                                              & 11                   & 368.45              & 0.27                    & 0.36                   & 1.18                 \\
Database                                               & 4                    & 332                 & 1                       & 0.25                   & 3.25                 \\
Logging and Monitoring                                 & 7                    & 256.57              & 0.86                    & 0.57                   & 4                    \\
Staging and Production                                 & 4                    & 238.75              & 1                       & 0.5                    & 1.75                 \\ \hline
Overall                                                & 10.09                & 725.79              & 0.97                    & 0.95                   & 3                    \\ \hline
\end{tabular}
\end{table*}

\begin{table*}[t]
\caption{Difficulty of architectural design issue groups.}
\label{difficult}
\centering
\addtolength{\tabcolsep}{-9pt}
\begin{tabular}{lcccc}
\hline
\textbf{Architectural Design Issue}      & \textbf{\# of Posts} & \textbf{\begin{tabular}[c]{@{}c@{}}\% of questions with no \\accepted answer\end{tabular}} & \textbf{Span (hours)} & \textbf{\begin{tabular}[c]{@{}c@{}}PD\\ =\\ (Avg. Ans / Avg. Views) * 100\end{tabular}} \\ \hline
Deployment                  &       6      & 100                        & -                     & 0.24                                                                                            \\
Security           &        11              & 72.73                      & 10.66                 & 0.31                                                                                            \\
Test                  &     5              & 60                         & 1.48                  & 0.14                                                                                            \\
Configuration             &     29          & 58.62                      & 58.27                 & 0.15                                                                                            \\
The Complexity of (Micro) Services at the Design Level       &  20               & 50                         & 17.75                 & 0.36                                                                                            \\
Setup and Architect DevOps Pipelines  & 6   & 50                         & 1.02                  & 0.16                                                                                            \\
Logging and Monitoring        &     7       & 42.86                      & 398.6                 & 0.56                                                                                            \\
Container        &        11                & 36.36                      & 1259.49               & 0.30                                                                                            \\
Availability and Scalability of Services    &  8    & 25                         & 182.26                & 0.07                                                                                            \\
Database                   &     4              & 25                         & 27.37                 & 0.53                                                                                            \\
Staging and Production      &       4             & 25                         & 0.62                  & 0.63                                                                                            \\ \hline
Overall   &      10.09            & 49.60                      & 195.75                & 0.31                                                                                            \\ \hline
\end{tabular}
\end{table*}

\subsubsection{Group 2: The Complexity of (Micro) Services at the Design Level}
Tables \ref{popular} and \ref{difficult} show that this group is the fifth most popular and difficult architectural design issue. This category includes 20 posts and is mainly about organizing several microservices appropriately at the design time, with which handling them would be much easier (e.g., quickly detecting bugs). The question ID \href{https://stackoverflow.com/questions/38440876}{38440876} is the highest viewed question in this category.

\begin{formal}\faComment
\textit{ \say{[...] I am using docker for the managing these services. With Docker-Compose I am able to define multiple services, but by default all the services defined in docker-compose.yml start. Is there any workaround for this? Also, is this a terrible strategy 
[but all my services use Docker, so is there an alternate way]?}} \footnotemark
\end{formal}\footnotetext{\url{https://stackoverflow.com/questions/38440876}}

A user who asked the most favorite question in this design issue group sought recommendations to ensure the compatibility of new microservices with existing standards in a given organization: 

\begin{formal}\faComment
\textit{ \say{How can I ensure that new microservices conform to organisation standards and include bug fixes/features from existing projects in a way that$'$s easy and intuitive for developers who want to start new projects? What are some best practises around resolving these issues?}} \footnotemark
\end{formal}\footnotetext{\url{https://devops.stackexchange.com/questions/2320}}

Another user was concern about synchronizing microservices versions in a DevOps pipeline: 

\begin{formal}\faComment
\textit{ \say{[...] I understand the DevOps lifecycle of a single monolithic web application, but cannot workout how a DevOps pipeline would work for microservices. [...] How do I keep the versions in sync?  Let's say the tools microservice uses blog version 2.3. But blog just got pushed to version 2.4, which is incompatible with tools.}} \footnotemark
\end{formal}\footnotetext{\url{https://stackoverflow.com/questions/56696750}\label{q2}}

A user advised that microservices should be backward compatible to avoid this issue: 

\begin{formal}\faComment
\textit{ \say{You need to be backwards compatible. Means if your blogs 2.4 version is not compatible with tools version 2.3 you will have high dependency and coupling which is going again one of the key benefits of micro-services [...] You can introduce a versioning system to your micro-services. If you have a breaking change to lets say an API you need to support the old version for some time still and create a new v2 of the new api.}}\footref{q2}.
\end{formal}

\subsubsection{Group 3: Security}
Developers concern about the security of tools and technologies (e.g., CI/CD servers), the production services, authentication of users, and authorization of microservices (i.e., 11 posts). A user asked a question on DevOps Stack Exchange: 

\begin{formal}\faComment
\textit{ \say{In my team we have one Jenkins server with one Master node \& one Slave node in the same server [...]
In a colleague$'$s team I have seen that, for some reason, they have one Master node and no slave node. All their builds run in Master node. Is this a bad practice? Are there any bad consequences for having only one Master node?}} \footnotemark
\end{formal}\footnotetext{\href{https://devops.stackexchange.com/questions/2105}{https://devops.stackexchange.com/questions/2105\label{q1}}}

A user with the highest reputation score in DevOps Stack Exchange responded to this question and warned that the proposed practice leads to a security compromise: 

\begin{formal}\faComment
\textit{ \say{In summary, from a security perspective it is a bad practice to run the jobs on the Master. Running the jobs on the Master nodes means that the jobs have unrestricted access into the JENKINS\_HOME directory.}} \footref{q1}
\end{formal}

Another user asked a question with 2000 views on Stack Overflow about the authentication of users in Azure DevOps:

\begin{formal}\faComment
\textit{ \say{I got a HTTP triggered azure function [...] I want to implement authentication to the azure function in such a way that only signed up user would able access the azure function through my web site. I could see many built-in authentications like Azure Functions, OAuth using Azure AD and other identity providers etc. I am looking for to way to authenticate users signed-up through my website, not with identity providers.
One solution I can think of is while signing up a register that user to Azure AD. Then while calling the API pass user credentials to the API and validate against AD. Can somebody please advice this is a good solution? [...] I don$'$t want to use any external auth provider.}} \footnotemark
\end{formal}\footnotetext{\url{https://stackoverflow.com/questions/55281504}}

Table \ref{difficult} shows that architectural security concern is the second most challenging concern as 72.73\% of the questions in this category have not yet had an accepted answer. As shown in Table \ref{popular}, on average, the questions of this category got 462.09 views, indicating that this architectural design issue has moderate popularity.

\subsubsection{Group 4: Container}
Eleven posts discuss how to containerize applications and allocate container resources. The questions in this category take the most time (on average, 1259.49 hours) to get an accepted answer (See Table \ref{difficult}). 
 The highest viewed question with a 25.3k reputation score on Stack Overflow is about how Docker can support the process of container resources allocation: 

\begin{formal}\faComment
\textit{ \say{[...] I've tried all these containers and them interactive with each other [...] I am going to run each one of these containers on a separate server. Is there something Docker provides to quickly/easily do this? Or should I set up docker on each server, copy relevant images to that server, and then start Docker-images manually on all servers?}} \footnotemark
\end{formal}\footnotetext{\url{https://stackoverflow.com/questions/34938674}}

The highest scored question on DevOps Stack Exchange seeks recommendations to deal with user interfaces inside containers:

\begin{formal}\faComment
\textit{ \say{Imagine in your stack you have RESTful services which also provide some rudimentary frontends, mostly for admin/other technical user use.
Do you include the user interface inside the container or are these two containers? Why?}} \footnotemark
\end{formal}\footnotetext{\url{https://devops.stackexchange.com/questions/2157}}

\subsubsection{Group 5: Availability and Scalability of Services}

Developers concern to achieve an acceptable level of scalability and availability of (micro) services being rapidly deployed on a (cloud) hosting like Amazon Web Services. As shown in Figure \ref{fig:number_of_posts}, this architectural design issue group include 8 posts. Table \ref{popular} shows that this group is the most popular architectural design issue (on average, 2280.37 views, 2.37 favorites, and 7.12 scores).
A user with a 9,892 reputation score asked the following question in DevOps Stack Exchange. This question is the most viewed (i.e., 16000 views), the most favorite (i.e., 17 favorites), and has the highest score (i.e., 40 scores) in this group.

\begin{formal}\faComment
\textit{ \say{How can the compound availability of the two systems above be calculated and documented for the business, potentially requiring rearchitecting if the business desires a higher service level than the architecture is capable of providing?}} \footnotemark
\end{formal}\footnotetext{\url{https://devops.stackexchange.com/questions/711}}

The question (ID \href{https://devops.stackexchange.com/questions/2779/availability-calculation-of-a-azure-service-fabric-stateful-application}{2779}) \textit{ \say{availability calculation of a Azure Service Fabric stateful application}} \footnote{\url{https://devops.stackexchange.com/questions/2779}} is apparently a duplicate of this question.  
The developers needed to calculate the availability of the services to replicate them if required. The scalability of services inside Docker containers was the concern of a user: 

\begin{formal}\faComment
\textit{ \say{What about scalability and high availability without using Docker? Can I just replicate my services and run them all at once in different hosts using a load balancer? And what about the scalability and availability of using Docker? Should I just put all my services inside containers and manage them with something like Kubernetes to achieve high availability?}} \footnotemark
\end{formal}\footnotetext{\url{https://stackoverflow.com/questions/48134800}}

\subsubsection{Group 6: Logging and Monitoring}
This group received the second-lowest popularity in Table \ref{popular}. This architectural design issue is about monitoring and tracking different instances of an application or a service in DevOps pipelines. For example, a user of DevOps Stack Exchange asked a question to find the best logging practice for logging the instances of a task: 
\begin{formal}\faComment
\textit{ \say{I have the following setting:
Create multiple workers, do a computation and terminate them after the computation is done.
So, every-time it$’$ll be a different instance running the task, so each host will have its own a log file, this will result in a huge list of files.
Is it a good practice? If not, what would be a better way for logging the task processing in this particular use-case?}} 
\end{formal}

Another user inquired about possible solutions for monitoring a microservice in the Docker container: 

\begin{formal}\faComment
\textit{ \say{This article about "How healthy is your Dockerized application?" explains the trouble with monitoring, but it doesn$'$t provide any good examples of how to actually monitor a microservice inside of the Docker container.}} \footnotemark
\end{formal}\footnotetext{\href{https://devops.stackexchange.com/questions/1195}{https://devops.stackexchange.com/questions/1195}\label{q3}}

The user used:

\begin{formal}\faComment
\textit{ \say{PM2 monit to monitor microservices but as user puts microservices into Docker containers, [he/she] lost the ability to access this data within a screen for all the various microservices which each run in their own Docker container.}} \footref{q3}
\end{formal}

Also, he/she examined:

\begin{formal}\faComment
\textit{ \say{The Docker Swarm monitoring, but this gave information about the state of the containers, not the microservice running inside of them.}} \footref{q3} 
\end{formal}

\subsubsection{Group 7: Deployment}
Table \ref{difficult} shows that developers have faced the most difficulty in addressing deployment issues as none of the questions in this category has had an accepted answer yet. Based on our analysis, the average number of views of the questions in this design issue is 420.5, indicating moderate popularity.
Developers mainly concern about how to deploy (micro) services in different stages of DevOps pipelines and make sure consistent updates among services: 

\begin{formal}\faComment
\textit{ \say{I was wondering how information on deployed applications are kept in real, production environments (yeah, I am a novice). For example, in a microservices based set up, if 5 microservices-based applications were deployed in production environments. Let$'$s say each of those had 4 - 6 instances each, are there special software for keeping record of these applications? I know the records can be easily written into a database but I can imagine some better solutions out there.}} \footnotemark
\end{formal}\footnotetext{\url{https://devops.stackexchange.com/questions/8952}}

\subsubsection{Group 8: Setup and Architect DevOps Pipelines}

This design issue includes questions that seek solutions and recommendations on organizing a DevOps pipeline and selecting the right tools among a wide range of available tools. This is the third most popular architectural design issue with a 1027 average number of views and 1.33 average number of favorites. 6 out of 100 posts belong to this architectural design issue group. The question ID \href{https://devops.stackexchange.com/questions/2542}{2542} is the most popular in this group with 5000 views, 5 favorites, and 7 scores:

\begin{formal}\faComment
\textit{ \say{Assume I put all the services in the same Project (GitLab repository are called Projects right?), can I then get the CI/CD system to build each component independently?}} \footnotemark
\end{formal}\footnotetext{\href{https://devops.stackexchange.com/questions/2542}{https://devops.stackexchange.com/questions/2542}\label{q2}}
The difficulty of this design issue is moderate, as only 50\% of the questions have the accepted answer, and it takes on average 1.02 hours to receive an accepted answer.

\subsubsection{Group 9: Test}

This group of architectural design issues includes 5 posts (Figure \ref{fig:number_of_posts}), asking how to test microservices developed and deployed through a DevOps pipeline effectively. As shown in Table \ref{popular}, this group is the second most popular architectural design issue among the users of Stack Overflow and DevOps Stack Exchange. Moreover, Table \ref{difficult} shows that achieving testable microservices systems in the DevOps context is the third most difficult architectural design issue. The most viewed question in this group is: 

\begin{formal}\faComment
\textit{ \say{Assume I commit a change to one service, and want to make sure it works with my other services before I deploy, how do I best setup integration tests that condition the success of my commit on the interaction with my other services?}} \footref{q2}
\end{formal}

And the highest scored question is:

\begin{formal}\faComment
\textit{ \say{I can do test per-server, which is good for some cases. But the problem is the following: my microservice architecture has a auto-discovery service, so some services are known after querying it. Is there any project to express this? I know I can tool around with ruby (or python if I choose testinfra or other).}} \footnotemark
\end{formal}\footnotetext{\url{https://devops.stackexchange.com/questions/731}}

\subsubsection{Group 10: Database}
We found 4 posts that seek solutions for database-related concerns, such as updating and deploying databases in different environments (e.g., the production database). 
The highest viewed question asked on Stack Overflow is:

\begin{formal}\faComment
\textit{ \say{I need to modify the Master and the Replica both to use the SSD storage type (GP2) with more space (2 TiB) and this has to be done via CloudFormation templates as the RDS instances mentioned above are part of a stack. But the issue is that these are production databases and a long outage [more than 2 hours] is not an option.
The storage size change alone is okay but the type change from magnetic to SSD is something of a grey area. There is no way (at least I know of) that I can be sure that it will be done in 2+ hours or how much time will be needed. I want to ask the community about the best practice here or if anyone has done this before with any work arounds (manual may be) that wouldn$'$t make the DBs out of sync of CloudFormation stack as well (like making a new replica manually with desired specs and promoting it to master for example)?}} \footnotemark
\end{formal} \footnotetext{\url{https://stackoverflow.com/questions/54535981}}

The second highest viewed question in DevOps Stack Exchange asks for suggestions to deal with the dependency between database instances and services: 

\begin{formal}\faComment
\textit{ \say{Due to a combination of business/enterprise requirements and our architect$'$s preferences we have arrived at a particular architecture that seems a bit off to me [...] the thing that rubs me the wrong way is the read-only DB instances spinning up inside the Docker containers for the services. The service itself and the DB would have drastically different demands in terms of load, so it would make a lot more sense if we could load balance them separately. I believe MySQL has a way of doing that with Master/Slave configurations, where new Slaves can be spun up whenever load gets high. Especially while we have our system in the cloud and are paying for each instance, spinning up a new instance of the whole service when we only need another DB instance seems wasteful (as does the opposite, spinning up a new DB copy when we really just need a web service instance) [...]
My largest concern is around the MS SQL Server implementation. Coupling the read only instances so tightly to the services feels wrong. Is there a better way to do this?}} \footnotemark
\end{formal}\footnotetext{\url{https://devops.stackexchange.com/questions/4058}}

\subsubsection{Group 11: Staging and Production}
The last category of architectural design issues includes questions that seek suggestions on organizing and configuring staging and production environments.
Popularity and difficulty of the questions in this category were the least compared to the questions in other categories (See Table \ref{popular} and \ref{difficult}).
A question with the most views, favorites, scores, and answers asked how to manage multiple staging environments during the migration from a monolith to a microservices architecture:

\begin{formal}\faComment
\textit{ \say{My company have recently embark on a platform architecture change from monolithic to microservice architecture. The entire migration might take years so as of now, we would still need to maintain the current monolithic application while slowly dismantling the application [...] Each team have their own stagings to manage this thus we have multiple stagings environment (11 in total) each with its own set of the legacy monolithic application. While transitioning to the microservice architecture [...] we would need to maintain all those stagings which that means we would need to have a copy of the microservice in each staging environment [...] Preferable if there are any answers in a different direction so we can have more options and variations to consider the pros and cons [...] The issue we face is as the number of microservices grow, that would take up a lot of server resources (we decided to use on premise server to host our Kubernetes and Proxmox Virtual Machine for the legacy monolith). Is there any infrastructure architecture that would reduce the resources that is required for this?}}\footnotemark
\end{formal}\footnotetext{\url{https://stackoverflow.com/questions/54837671}}

As shown in Table \ref{difficult}, while the questions in this group got the least accepted answer (i.e., only 25\% of the questions have an accepted answer), the questions waited the least time to receive an accepted answer.

\section{Discussion}\label{discussionSec}
In the following, we discuss some of the main findings from our study, and then we present implications of our results, which are expected to be helpful for researchers, organizations, and practitioners.

\subsection{Summary of Main Findings}
\subsubsection{The Qualitative Case Study (RQ1)}
This study identified 8 architectural design issues that two teams (TeamA and TeamB) from a company faced during their DevOps transformation, along with an in-depth analysis of the consequences of decisions made to address these 8 architectural design issues. TeamA decided to use one repository to manage all software modules and artifacts at the development side and had one deployment unit in production. On the other hand, TeamB chose to use one repository for each module and artifact of its system and release the system as multiple deployment units.  Despite this difference, the decisions made by both teams mainly aimed to improve deployability, testability, supportability, and modifiability.
\subsubsection{The Content Analysis Study (RQ2)}
This study found that software practitioners discussed and reported 11 groups of architectural design issues in the context of DevOps in Stack Overflow and DevOps Stack Exchange. Among these 11 groups, \say{Configuration} and \say{Complexity of (Micro) services at the Design Level} have the most discussions (i.e., they have the most number of posts). However, we found that the top three most popular architectural design issue groups are \say{Availability and Scalability of Services}, \say{Test}, and \say{Setup and Architect DevOps Pipelines} and the top three most difficult architectural design issue groups are \say{Deployment}, \say{Security}, and \say{Test}. Concerning popularity, on average, the questions in the architectural design issue groups have 725.79 views, 0.97 favorites, 0.95 comments, and 3 scores. In terms of difficulty, it takes, on average, 195.75 hours that a question gets an accepted answer.
\subsubsection{Cross-Analysis}

As shown in Table \ref{tbl:crossanalysis}, all design issues identified from the qualitative case study, except for Design Issue 8, can be linked to one or more architectural design issue groups discovered in the content analysis study. For example, Design Issue 1 (different instances of an application/service should be treated as the same artifact in different environments) includes discussions around configuration, deployment, and production environment (See Section \ref{sec:DIssue1}) . In another example, the teams from the case company and developers in Stack Overflow and DevOps Stack Exchange aimed to enhance the monitorability and loggability of applications in DevOps (Design Issue 5 and Group 6 in Table \ref{tbl:crossanalysis}). Only one design issue (Design Issue 8: teams should be reorganized to effectively adopt DevOps) faced by the teams has no corresponding issue in the content analysis study. This is partly because developers do not usually discuss team structures in online community websites. Further to this, we did not find that TeamA and TeamB (frequently) discussed about security, container, and database issues. All this shows that although the design issues in the case company are specific and contextual, other software practitioners also discuss similar design issues in Stack Overflow and DevOps Stack Exchange. Further, this shows that both studies complement each other.

\begin{table*}[!h]
    \centering
    \includegraphics[scale=0.63]{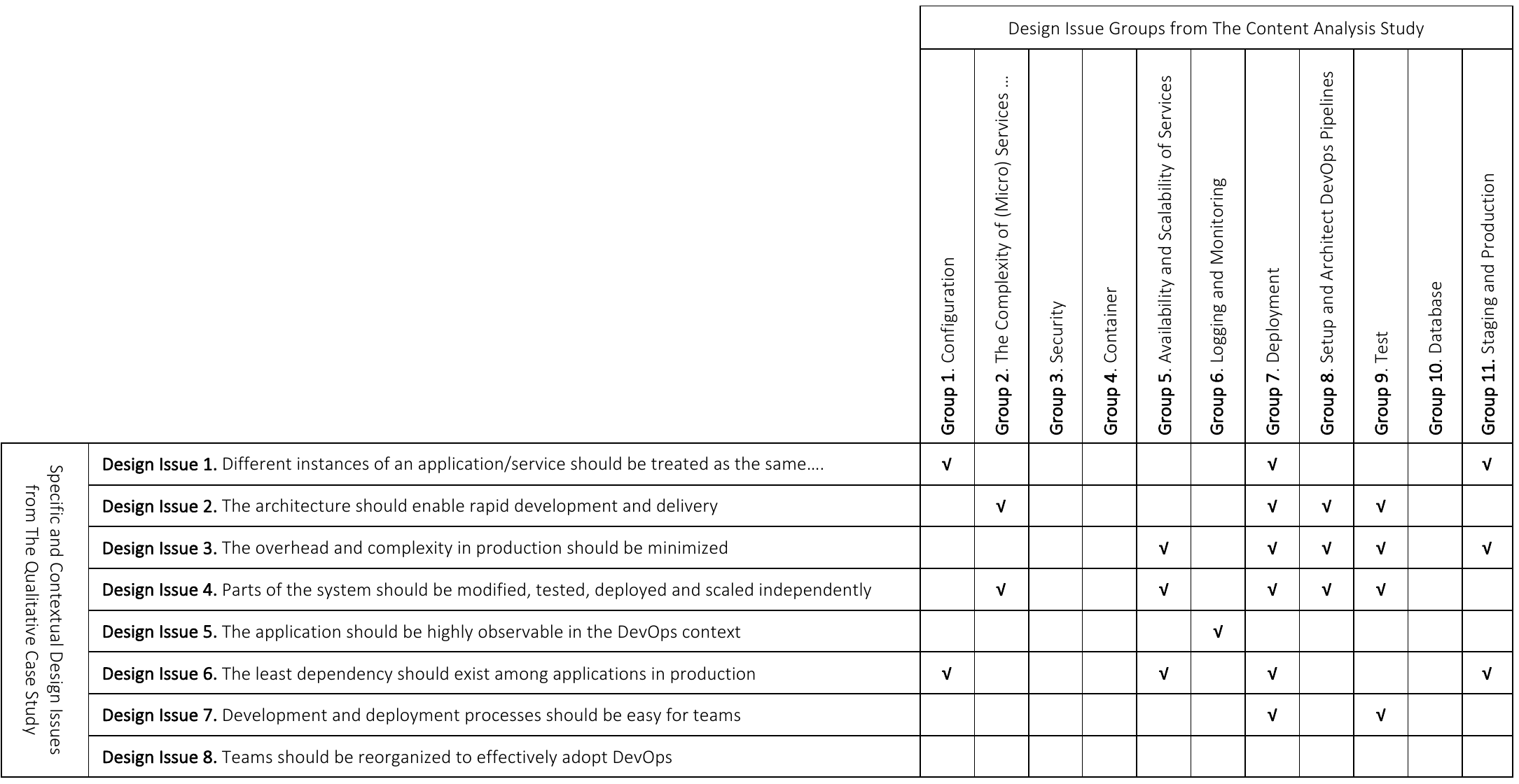}
    \caption{A cross-analysis of designs issues found in the qualitative case study and the content analysis study}
    \label{tbl:crossanalysis}
\end{table*}

\subsection{Implications for Researchers and Practitioners}
\subsubsection{Characteristics of DevOps-driven Architectures}

A growing number of organizations have started or plan to leverage the microservices architecture style as a driver to succeed in DevOps \cite{balalaie2016microservices,callanan2016devops,waseem2020systematic}. Our study shows that 28 out of 100 (28\%) questions asked in Stack Overflow and DevOps Stack Exchange are exclusively about (micro)services-based systems in the DevOps context: how to reduce the complexity of (micro) services at the design level and how to achieve an acceptable level of scalability and availability of (micro)services. Among two studied teams in the qualitative case study, TeamA started with the microservices architecture style, but they finally decided to have one monolithic deployment unit in the production as they faced difficulties in deploying the initial microservices system. On the other hand, while TeamB chose not to follow TeamA's deployment approach (i.e., one deployment unit in production), they were reluctant to fully embrace the microservices architecture style. While TeamA and TeamB realized that the fundamental limitation to rapid and safe delivery resides in their systems’ architectures, they found that loosely coupled architecture and prioritizing deployability, testability, supportability, and modifiability over other quality attributes can guarantee to a large extent their success in DevOps (i.e., confirming and extending \cite{shahin2019empirical,chen2015towards}). This is because improving/addressing these quality attributes was the target of the majority of the identified decisions. For example, deployability was positively influenced by six decisions made to address Design Issues 1, 2, 3, 4, 6, and 7. Our findings are in line with the 2017 State of DevOps Report \cite{forsgren20172017}, which revealed that apart from architecting a system with microservices style, service-oriented architecture or any other architecture styles, loosely coupled architectures and teams are the most significant contributors to DevOps success (i.e., releasing software changes at high velocity). Etsy is a notable example of this scenario, as it has successfully implemented DevOps using a monolithic system \cite{schauenberg2014development}. Here monolith means a single build system, in which all the functionality is managed and deployed in one deployable unit \cite{Self-Contained,dragoni2017microservices}. This does not necessarily mean that a monolith is a highly coupled architecture.

We observed that TeamB is more likely to have loosely coupled architecture than that of TeamA. It is mainly because changing the architecture of TeamA’s platform is not straightforward, as TeamA’s architect needed to get involved frequently whenever the architectural changes required. Conversely, TeamB members could apply small-scale changes to their architecture without the need for their architect. However, in both systems applying large-scale changes required the involvement of their architects. Another sign that shows the architecture of TeamB’s platform is more modular is that TeamB had higher deployment frequency to lower environments (e.g., test environment) compared to TeamA (See Table \ref{diffenv}). Other studies \cite{shahin2019empirical,erder2015continuous} reveal that the architectures in the DevOps context should support evolutionary changes. This implies that architectural decisions should be delayed until necessary (i.e., delaying decisions to the last possible moment) \cite{erder2015continuous}. However, it does not mean that there is no longer a need to have a long-term vision of architecture. Put another way, core architectural decisions need to be made at the early stage of development in the DevOps context. By ignoring this necessity, TeamA experienced a pain point in the architecting process as they ended up major refactoring of the whole stack of their system several times.

\begin{tcolorbox}[colback=gray!2!white,colframe=white!80!black]

\textbf{Implication.} We argue that DevOps-driven architecture can be characterized by being loosely coupled and prioritizing deployability, testability, supportability, and modifiability.  
\end{tcolorbox}
\subsubsection{Challenges Relating to Operations Tasks}
Our findings in the qualitative case study show that establishing cross-functional, autonomous teams was one of the key decisions made in the case company to implement DevOps (i.e., the decision made for Decision Issue 8). We observed that operations expert is not embodied as a distinct role on both TeamA and TeamB, or there is no dedicated operations team in the case company. The operations skills are regarded as a skillset that blends with other skills such as software development, and the operations tasks need to be performed by all members. The content analysis study on Stack Overflow and DevOps Stack Exchange also shows that developers face difficulties in working with, setting up, and configuring tools (e.g., \textit{Ansible} (\faWrench{})) used in the operations side. Almost 60\% of the posts concerning configuring DevOps tools and technologies (i.e., Design Issue Group 1 in the content analysis study) have no accepted answer. 
\begin{tcolorbox}[colback=gray!2!white,colframe=white!80!black]
\textbf{Implication.} Although DevOps encourages developers to do operations tasks, we argue that performing the operations tasks that require deep expertise in operations (e.g., writing Ansible playbook) is burdensome for developers.
\end{tcolorbox}
\subsubsection{Operations Experts in Software Development Team}
The complexity of some operations tasks may cause a significant part of developers' time to be spent on the operations tasks rather than the development tasks (i.e., the negative consequence of the decisions made for Design Issues 6 and 8 on team effort). We also found that this can give developers a good excuse to not perform the operations tasks optimally. This can be exemplified by the Ansible playbooks written by the members of TeamA and TeamB as part of their new responsibilities in DevOps. While both teams’ members found that the quality of the written Ansible playbooks is not good enough, we found that there is not any demand to improve them. In our view, this could stem from the fact that operations responsibilities are ambiguous for the software development team, and some of the operations tasks are not clear who should do. Furthermore, in the DevOps transformation, the development side is more emphasized than the operations side \cite{shahin2019empirical}. This could be mainly justified by the fact that most of the business values come from the development side (e.g., adding more features). Embedding operations specialists in software development teams for complex operations tasks would benefit people from the development side as they can concentrate more on real software engineering tasks \cite{Serving,DevOps}. This is in line with \cite{shahin2017adopting}, showing that most surveyed organizations (76 out of 93) prefer having a distinct operations team or operations experts for specific operations tasks.
\begin{tcolorbox}[colback=gray!2!white,colframe=white!80!black]
\textbf{Implication.} We emphasize that operations specialists need to be embedded in software development teams for complex operations tasks. Further to this, given the difficulties developers face in using and configuring operations side tools, we emphasize that tool vendors consider applying standards and mechanisms that reduce the difficulty and effort required for using and learning such tools.

\end {tcolorbox}

\subsubsection{Investment in (Automated) Testing}

Many architectural decisions (i.e., decisions made to address Design Issues 2, 3, 4, 7) made by the two teams from the case company (Section \ref{sectionStudy1}) have a positive consequence on the testability of an application in the DevOps context. The \say{Test} design issue group is the second most popular design issue in DevOps raised by Stack Overflow and DevOps Stack Exchange users (See Section \ref{sectionSecondStudy}). Furthermore, Stack Overflow and DevOps Stack Exchange questions were asked about test-related issues are the third-most difficult questions to receive an accepted answer. In general, our study reveals the following problems concerning testing applications in the context of DevOps: lack of having good test coverage, lack of comprehensive end-to-end integration test automation, and conducting performance, security, and acceptance tests out of DevOps pipelines. 

Moreover, in both studies, we found that developers do not want or cannot automate some sorts of tests (e.g., acceptance test) in DevOps pipelines. For instance, as the overhead of maintaining Selenium-based user interface testing increased (e.g., because user interface changes a lot), TeamA found that it is better to turn off user interface testing in the deployment pipeline. That is why currently, they do it manually on the release branch. While having an automated deployment to production and designing with deployability in mind can help, to some extent, software development teams increase the deployment speed, we observed that test automation has a strong influence on the deployment speed. This, in turn, enables organizations to adopt continuous delivery or deployment as one of the most emphasized DevOps practices \cite{bass2017software}. Despite this, we acknowledge that truly implementing continuous delivery and deployment might be influenced by other factors such as socio-technical factors (e.g., the circumstance of the studied projects in the case company also does not allow the teams to deploy the changes multiple times a day). 
\begin{tcolorbox}[colback=gray!2!white,colframe=white!80!black]

\textbf{Implication}: We argue that organizations should ensure to have good test coverage, write tests that less consume cycle time of a DevOps pipeline, and automate tests (e.g., performance) that occur during the last stages of DevOps pipelines for successfully implementing continuous delivery and deployment \cite{shahin2017continuous,bass2017software}. This provides confidence to deploy to production continuously and automatically.
    
\end{tcolorbox}

\section{Threats to Validity}\label{secThreats}
Our mixed-methods study may have a few limitations, which we discuss below from the qualitative research perspective \cite{guba1981criteria,stol2014key}.

\subsection{Transferability}
Generalizability is a concern in any qualitative research. We have tried to minimize this limitation by conducting two complementary studies. 
The case study investigated two independent teams from one organization working on two different projects to understand the specific and contextual architectural design issues they faced during DevOps transformation. Hence, the design challenges identified from the case study may not be generalized to other teams or organizations (e.g., teams working on different domains). We tried to minimize this threat by conducting a content analysis of 100 posts from Stack Overflow and DevOps Stack Exchange. While the content analysis study provides a broader overview of possible architectural design issues in DevOps from a more general perspective, developers may discuss other important architectural design challenges in other online community websites and forums (e.g., GitHub, Reddit). Hence, other developer community websites should be investigated, and more case studies need to be conducted to identify the most comprehensive set of architectural design challenges in DevOps. 
\subsection{Credibility}
Using three different data sources (i.e., interview, documentation, and question and answer websites) and investigating two different teams ensured that the obtained findings, to a large extent, are plausible. The selection of the participants in the case study can have brought a threat to the credibility of our findings. To recruit motivated participants, we ensured that personal details, opinions, and thoughts would be regarded as strictly confidential and would not be divulged to the researchers and other team members by making a confidential disclosure agreement. After discussing the objectives of the case study with the CTO and two key informants at the case company, the CTO introduced us to a few suitable persons from each team for the interviews who were invited for participation. This gives us confidence that the team members who chose to participate were likely more willing and had the right types of competencies to provide an unbiased opinion. The number of interviews (i.e., six interviews) is relatively low, which can be a threat. We tried to reduce this threat by analyzing a set of documents (e.g., team discussions forum) provided by the teams. Moreover, three of the interviewees had almost the same job titles, i.e., two architects (solution, software) in TeamA and one system architect in TeamB. It should be noted that those who held these job titles did coding, testing, and operations tasks in their respective teams. Another possible limitation comes from the interview questions. Our interview questions were primarily designed based on a comprehensive systematic review \cite{shahin2017continuous} and the existing empirical studies on software architecture and DevOps \cite{forsgren20172017,shahin2019empirical,chen2015towards}. In addition, we tried to fine-tune the questions and asked appropriate follow-up questions according to the participants’ responses and their projects. 

The credibility of our findings in the content analysis study could be threatened by the selection process of the posts. While we only selected posts with \say{architect}, \say{DevOps}, and \say{microservices} tags, design issues in DevOps may also be expressed in posts with other tags, such as posts with \say{continuous delivery} and \say{design} tags. Further to this, there might be posts that discuss software architecture in DevOps but do not use \say{architect}, \say{DevOps}, and \say{microservices} tags. Hence, our results may not provide a complete view of all Stack Overflow and DevOps Stack Exchange users regarding design issues in DevOps. Some of the architectural design issue groups include a small number of posts. For example, \say{Database} and \say{Stating and Production} groups have 4 posts each. We acknowledge that this can be another threat to the credibility of our findings.

Another threat to the credibility of our study is whether the architectural design issues identified from the case study and content analysis study (and their corresponding decisions in the case study) were exclusively driven by DevOps, not because of other confounding factors. We accept that there might be decisions and design issues that are not strictly related to DevOps or could be found in other software development methodologies (e.g., the decision 1 made for Design Issue 1 in the case study is like one of best practices in twelve-factor methodology \cite{Twelve-Factor} or migrating applications to the cloud \cite{cito2015making}). To mitigate this threat in the case study, we took two actions. In the informal meeting with CTO and two informants, we described the objective of our study (e.g., identifying architectural design issues and their corresponding architectural decisions in DevOps). During the interviews, we also encouraged the interviewees to mainly discuss the changes in the architecting process in the context of DevOps. In the content analysis study, a post with \say{DevOps}, \say{architect}, and \say{microservices} tags does not necessarily discuss architecture design issues in DevOps. Our strategy to detect such posts was to thoroughly read and analyze the question title, question body, question's comments, answers, and answer's comments of all identified posts. We removed 32 posts with such tags that did not discuss design issues relevant to DevOps.



\subsection{Confirmability}
In both the case study and content analysis study, data analysis was conducted by one person. While this helped to obtain consistency in the results \cite{tomasdottir2017and}, it can be a potential threat to the validity of the findings. In both studies, this threat was mitigated to some extent by organizing several internal discussions between the authors to review and verify the findings and solicit feedback. In the case study, we further minimized the subjective bias by asking the CTO of the case company to review and provide feedback on the early version of the findings of the case study. The CTO mainly reviewed the identified design issues and their corresponding decisions and checked if our findings breach the signed non-disclosure agreement. Furthermore, other research \cite{stol2014key} highlights that data triangulation strategy, which previously described for improving credibility, can be used to establish confirmability as it can reduce the subjectivity of the researcher’s understanding and judgment. 


\section{Conclusion}\label{secConclusion}
In this paper, we conducted a mixed-methods study consisting of a qualitative case study complemented by a content analysis of 100 posts from Stack Overflow and DevOps Stack Exchange to learn about architectural design issues of interest in DevOps. The qualitative case study focused on two teams in a case company. It found eight architectural design issues (e.g., different instances of an application/service should be treated as the same in different environments.) the teams faced and their corresponding architectural decisions (e.g., external configuration) during DevOps transformation. The content analysis of 100 posts provided a more general perspective on architectural design issues in DevOps and classified them into eleven groups.
Among these eleven groups, \say{Configuration} and \say{The complexity of (Micro) Services at the Design Level} are the most frequently discussed, \say{Availability and Scalability of Services}, and \say{Test} are the most popular, and \say{Deployment}, \say{Security}, and \say{Test} the most challenging architectural design issues. 
Our findings suggest that DevOps success is best associated with loosely coupled architectures that prioritize deployability, testability, supportability, and modifiability over other quality attributes. Finally, our observations show that (1) operations specialists should be part of the software development teams to perform advanced and complex operations tasks, and (2) investment in automated testing (automating tests that occur during the last stages of DevOps pipelines) can largely guarantee the success of software organizations in release software changes continuously.

\section*{Acknowledgements} 
When some parts of this work were carried out, the first author was affiliated with the University of Adelaide and supported by the Australian Government Research Training Program Scholarship and Data61. We like to thank the participants and the CTO of the case company.

\section*{Data Availability Statement}
 
Research data are not shared.

\printendnotes

\bibliography{sample}

\begin{thebibliography}{73}
\providecommand{\natexlab}[1]{#1}
\providecommand{\url}[1]{\texttt{#1}}
\providecommand{\urlprefix}{}

\bibitem[{Forsgren et~al.(2017)Forsgren, N and Kim, G and Kersten, N and
  Humble, J and Brown, A}]{forsgren20172017}
Forsgren N, Kim G, Kersten N, Humble J, Brown A.
\newblock 2017 State of DevOps report.
\newblock https://puppet.com/resources/report/2017-state-devops-report/: Puppet
  and DORA; 2017.

\bibitem[{Shahin et~al.(2017)Shahin, Mojtaba and Babar, Muhammad Ali and Zhu,
  Liming}]{shahin2017continuous}
Shahin M, Babar MA, Zhu L.
\newblock Continuous integration, delivery and deployment: a systematic review
  on approaches, tools, challenges and practices.
\newblock IEEE Access 2017;5:3909--3943.

\bibitem[{Bass et~al.(2015)Bass, Len and Weber, Ingo and Zhu,
  Liming}]{bass2015devops}
Bass L, Weber I, Zhu L.
\newblock DevOps: A software architect's perspective.
\newblock Addison-Wesley Professional; 2015.

\bibitem[{Leite et~al.(2019)Leite, Leonardo and Rocha, Carla and Kon, Fabio and
  Milojicic, Dejan and Meirelles, Paulo}]{leite2019survey}
Leite L, Rocha C, Kon F, Milojicic D, Meirelles P.
\newblock A survey of DevOps concepts and challenges.
\newblock ACM Computing Surveys (CSUR) 2019;52(6):1--35.

\bibitem[{Rahman and Williams(2016)Rahman, Akond Ashfaque Ur and Williams,
  Laurie}]{rahman2016software}
Rahman AAU, Williams L.
\newblock Software security in devops: Synthesizing practitioners’
  perceptions and practices.
\newblock 2016 IEEE/ACM International Workshop on Continuous Software Evolution
  and Delivery (CSED), IEEE; 2016. p. 70--76.

\bibitem[{Jaatun et~al.(2017)Jaatun, Martin Gilje and Cruzes, Daniela S and
  Luna, Jesus}]{jaatun2017devops}
Jaatun MG, Cruzes DS, Luna J.
\newblock Devops for better software security in the cloud invited paper.
\newblock Proceedings of the 12th International Conference on Availability,
  Reliability and Security, ACM; 2017. p. 1--6.

\bibitem[{M{\"a}kinen et~al.(2016)M{\"a}kinen, Simo and Lepp{\"a}nen, Marko and
  Kilamo, Terhi and Mattila, Anna-Liisa and Laukkanen, Eero and Pagels, Max and
  M{\"a}nnist{\"o}, Tomi}]{makinen2016improving}
M{\"a}kinen S, Lepp{\"a}nen M, Kilamo T, Mattila AL, Laukkanen E, Pagels M,
  et~al.
\newblock Improving the delivery cycle: A multiple-case study of the toolchains
  in Finnish software intensive enterprises.
\newblock Information and Software Technology 2016;80:175--194.

\bibitem[{Kang et~al.(2016)Kang, Hui and Le, Michael and Tao,
  Shu}]{kang2016container}
Kang H, Le M, Tao S.
\newblock Container and microservice driven design for cloud infrastructure
  devops.
\newblock 2016 IEEE International Conference on Cloud Engineering (IC2E), IEEE;
  2016. p. 202--211.

\bibitem[{Wettinger et~al.(2016)Wettinger, Johannes and Breitenb{\"u}cher, Uwe
  and Kopp, Oliver and Leymann, Frank}]{wettinger2016streamlining}
Wettinger J, Breitenb{\"u}cher U, Kopp O, Leymann F; Elsevier.
\newblock Streamlining DevOps automation for Cloud applications using TOSCA as
  standardized metamodel.
\newblock Future Generation Computer Systems 2016;56:317--332.

\bibitem[{van Hoorn et~al.(2017)van Hoorn, Andre and Jamshidi, Pooyan and
  Leitner, Philipp and Weber, Ingo}]{van2017report}
van Hoorn A, Jamshidi P, Leitner P, Weber I.
\newblock Report from GI-Dagstuhl Seminar 16394: Software Performance
  Engineering in the DevOps World.
\newblock arXiv preprint arXiv:170908951 2017;.

\bibitem[{Lwakatare et~al.(2019)Lwakatare, Lucy Ellen and Kilamo, Terhi and
  Karvonen, Teemu and Sauvola, Tanja and Heikkil{\"a}, Ville and Itkonen, Juha
  and Kuvaja, Pasi and Mikkonen, Tommi and Oivo, Markku and Lassenius,
  Casper}]{lwakatare2019devops}
Lwakatare LE, Kilamo T, Karvonen T, Sauvola T, Heikkil{\"a} V, Itkonen J,
  et~al.; Elsevier.
\newblock DevOps in practice: A multiple case study of five companies.
\newblock Information and Software Technology 2019;114:217--230.

\bibitem[{Luz et~al.(2019)Luz, Welder Pinheiro and Pinto, Gustavo and
  Bonif{\'a}cio, Rodrigo}]{luz2019adopting}
Luz WP, Pinto G, Bonif{\'a}cio R.
\newblock Adopting DevOps in the real world: A theory, a model, and a case
  study.
\newblock Journal of Systems and Software 2019;157:110384.

\bibitem[{Erich et~al.(2017)Erich, FMA and Amrit, Chintan and Daneva,
  Maya}]{erich2017qualitative}
Erich F, Amrit C, Daneva M.
\newblock A qualitative study of DevOps usage in practice.
\newblock Journal of Software: Evolution and Process 2017;29(6):e1885.

\bibitem[{Lwakatare et~al.(2016)Lwakatare, Lucy Ellen and Karvonen, Teemu and
  Sauvola, Tanja and Kuvaja, Pasi and Olsson, Helena Holmstr{\"o}m and Bosch,
  Jan and Oivo, Markku}]{lwakatare2016towards}
Lwakatare LE, Karvonen T, Sauvola T, Kuvaja P, Olsson HH, Bosch J, et~al.
\newblock Towards DevOps in the embedded systems domain: Why is it so hard?
\newblock 2016 49th Hawaii International Conference on System Sciences (HICSS),
  IEEE; 2016. p. 5437--5446.

\bibitem[{Shahin et~al.(2017)Shahin, Mojtaba and Zahedi, Mansooreh and Babar,
  Muhammad Ali and Zhu, Liming}]{shahin2017adopting}
Shahin M, Zahedi M, Babar MA, Zhu L.
\newblock Adopting continuous delivery and deployment: Impacts on team
  structures, collaboration and responsibilities.
\newblock Proceedings of the 21st International Conference on Evaluation and
  Assessment in Software Engineering (EASE), ACM; 2017. p. 384--393.

\bibitem[{Nybom et~al.(2016)Nybom, Kristian and Smeds, Jens and Porres,
  Ivan}]{nybom2016impact}
Nybom K, Smeds J, Porres I.
\newblock On the impact of mixing responsibilities between devs and ops.
\newblock 2016 17th International Conference on Agile Software Development
  (XP), Springer; 2016. p. 131--143.

\bibitem[{Skelton and Pais(2019)Skelton, Matthew and Pais,
  Manuel}]{skelton2019team}
Skelton M, Pais M.
\newblock Team Topologies: Organizing Business and Technology Teams for Fast
  Flow.
\newblock IT Revolution; 2019.

\bibitem[{Leite et~al.(2020)Leite, Leonardo and Pinto, Gustavo and Kon, Fabio
  and Meirelles, Paulo}]{leite2020organization}
Leite L, Pinto G, Kon F, Meirelles P.
\newblock The Organization of Software Teams in the Quest for Continuous
  Delivery: A Grounded Theory Approach.
\newblock arXiv preprint arXiv:200808652 2020;.

\bibitem[{Shahin et~al.(2019)Shahin, Mojtaba and Zahedi, Mansooreh and Babar,
  Muhammad Ali and Zhu, Liming}]{shahin2019empirical}
Shahin M, Zahedi M, Babar MA, Zhu L.
\newblock An empirical study of architecting for continuous delivery and
  deployment.
\newblock Empirical Software Engineering 2019;24(3):1061--1108.

\bibitem[{Bellomo et~al.(2014)Bellomo, Stephany and Ernst, Neil and Nord,
  Robert and Kazman, Rick}]{bellomo2014toward}
Bellomo S, Ernst N, Nord R, Kazman R.
\newblock Toward design decisions to enable deployability: Empirical study of
  three projects reaching for the continuous delivery holy grail.
\newblock 2014 44th Annual IEEE/IFIP International Conference on Dependable
  Systems and Networks (DSN), IEEE; 2014. p. 702--707.

\bibitem[{Chen(2015)Chen, Lianping}]{chen2015towards}
Chen L.
\newblock Towards architecting for continuous delivery.
\newblock 2015 12th Working IEEE/IFIP Conference on Software Architecture
  (WICSA), IEEE; 2015. p. 131--134.

\bibitem[{Balalaie et~al.(2016)Balalaie, Armin and Heydarnoori, Abbas and
  Jamshidi, Pooyan}]{balalaie2016microservices}
Balalaie A, Heydarnoori A, Jamshidi P.
\newblock Microservices architecture enables devops: Migration to a
  cloud-native architecture.
\newblock IEEE Software 2016;33(3):42--52.

\bibitem[{Callanan and Spillane(2016)Callanan, Matt and Spillane,
  Alexandra}]{callanan2016devops}
Callanan M, Spillane A.
\newblock DevOps: making it easy to do the right thing.
\newblock IEEE Software 2016;33(3):53--59.

\bibitem[{Chen(2018)Chen, Lianping}]{chen2018microservices}
Chen L.
\newblock Microservices: Architecting for continuous delivery and devops.
\newblock 2018 IEEE International Conference on Software Architecture (ICSA),
  IEEE; 2018. p. 39--397.

\bibitem[{Zdun et~al.(2019)Zdun, Uwe and Wittern, Erik and Leitner,
  Philipp}]{zdun2019emerging}
Zdun U, Wittern E, Leitner P.
\newblock Emerging Trends, Challenges, and Experiences in DevOps and
  Microservice APIs.
\newblock IEEE Software 2019;37(1):87--91.

\bibitem[{Waseem et~al.(2020)Waseem, Muhammad and Liang, Peng and Shahin,
  Mojtaba}]{waseem2020systematic}
Waseem M, Liang P, Shahin M.
\newblock A systematic mapping study on microservices architecture in devops.
\newblock Journal of Systems and Software 2020;170:110798.

\bibitem[{XebiaLabs(????)}]{Exploring}
XebiaLabs.
\newblock Exploring Microservices: 14 Questions Answered By Experts.
\newblock https://digital.ai/resources/library: XebiaLabs;.

\bibitem[{Schmidt(2016)M. Schmidt}]{SchmidtDevOps}
Schmidt M, DevOps and Continuous Delivery: Not the Same.
\newblock Web site: [Last accessed: 11/06/2021]; 2016.
\newblock \url{https://devops.com/devops-and-continuous-delivery-not-same/}.

\bibitem[{Laukkarinen et~al.(2017)Laukkarinen, Teemu and Kuusinen, Kati and
  Mikkonen, Tommi}]{laukkarinen2017devops}
Laukkarinen T, Kuusinen K, Mikkonen T.
\newblock DevOps in regulated software development: case medical devices.
\newblock 2017 IEEE/ACM 39th International Conference on Software Engineering:
  New Ideas and Emerging Technologies Results Track (ICSE-NIER), IEEE; 2017. p.
  15--18.

\bibitem[{Shahin et~al.(2017)Shahin, Mojtaba and Babar, Muhammad Ali and
  Zahedi, Mansooreh and Zhu, Liming}]{shahin2017beyond}
Shahin M, Babar MA, Zahedi M, Zhu L.
\newblock Beyond continuous delivery: an empirical investigation of continuous
  deployment challenges.
\newblock 2017 ACM/IEEE International Symposium on Empirical Software
  Engineering and Measurement (ESEM), IEEE; 2017. p. 111--120.

\bibitem[{Taibi et~al.(2017)Taibi, Davide and Lenarduzzi, Valentina and Pahl,
  Claus and Janes, Andrea}]{taibi2017microservices}
Taibi D, Lenarduzzi V, Pahl C, Janes A.
\newblock Microservices in agile software development: a workshop-based study
  into issues, advantages, and disadvantages.
\newblock Proceedings of the XP2017 Scientific Workshops, ACM; 2017. p. 1--5.

\bibitem[{Newman and Fowler(2020)Newman, Sam and Fowler,
  Martin}]{whenuseMicroservices}
Newman S, Fowler M, When To Use Microservices (And When Not To!).
\newblock Web site: [Last accessed: 11/06/2021]; 2020.
\newblock \url{https://www.youtube.com/watch?v=GBTdnfD6s5Q}.

\bibitem[{Zhu et~al.(2016)Zhu, Liming and Bass, Len and Champlin-Scharff,
  George}]{zhu2016devops}
Zhu L, Bass L, Champlin-Scharff G.
\newblock DevOps and its practices.
\newblock IEEE Software 2016;33(3):32--34.

\bibitem[{Shahin and Babar(2020)Shahin, Mojtaba and Babar, M. Ali}]{shahin2020}
Shahin M, Babar MA.
\newblock On the Role of Software Architecture in DevOps Transformation: An
  Industrial Case Study.
\newblock 14th International Conference on Software and Systems Process
  (ICSSP), ACM; 2020. p. 175–184.
\newblock Https://doi.org/10.1145/3379177.3388891.

\bibitem[{M{\aa}rtensson et~al.(2017)M{\aa}rtensson, Torvald and St{\aa}hl,
  Daniel and Bosch, Jan}]{maartensson2017continuous}
M{\aa}rtensson T, St{\aa}hl D, Bosch J.
\newblock Continuous integration impediments in large-scale industry projects.
\newblock 2017 14th IEEE International Conference on Software Architecture
  (ICSA), IEEE; 2017. p. 169--178.

\bibitem[{Schermann et~al.(2016)Schermann, Gerald and Cito, J{\"u}rgen and
  Leitner, Philipp and Zdun, Uwe and Gall, Harald}]{schermann2016empirical}
Schermann G, Cito J, Leitner P, Zdun U, Gall H.
\newblock An empirical study on principles and practices of continuous delivery
  and deployment.
\newblock https://peerj.com/preprints/1889/: PeerJ Preprints; 2016.

\bibitem[{Di~Nitto et~al.(2016)Di Nitto, Elisabetta and Jamshidi, Pooyan and
  Guerriero, Michele and Spais, Ilias and Tamburri, Damian A}]{di2016software}
Di~Nitto E, Jamshidi P, Guerriero M, Spais I, Tamburri DA.
\newblock A software architecture framework for quality-aware DevOps.
\newblock Proceedings of the 2nd International Workshop on Quality-Aware DevOps
  (QUDOS), ACM; 2016. p. 12--17.

\bibitem[{Moses et~al.(2020)Moses, Openja and Bram, Adams and Foutse,
  Khomh}]{Moses2020Analysis}
Moses O, Bram A, Foutse K.
\newblock Analysis of Modern Release Engineering Topics – A Large-Scale Study
  using StackOverflow -.
\newblock Proceedings of the 36th IEEE International Conference on Software
  Maintenance and Evolution (ICSME), IEEE; 2020. p. 104--114.

\bibitem[{Capilla et~al.(2016)Capilla, Rafael and Jansen, Anton and Tang,
  Antony and Avgeriou, Paris and Babar, Muhammad Ali}]{capilla201610}
Capilla R, Jansen A, Tang A, Avgeriou P, Babar MA.
\newblock 10 years of software architecture knowledge management: Practice and
  future.
\newblock Journal of Systems and Software 2016;116:191--205.

\bibitem[{Oliveira et~al.(2018)Oliveira, Nigini and Muller, Michael and
  Andrade, Nazareno and Reinecke, Katharina}]{oliveira2018exchange}
Oliveira N, Muller M, Andrade N, Reinecke K.
\newblock The exchange in StackExchange: Divergences between Stack Overflow and
  its culturally diverse participants.
\newblock Proceedings of the ACM on Human-Computer Interaction
  2018;2(CSCW):1--22.

\bibitem[{Vadlamani and Baysal(2020)Vadlamani, Sri Lakshmi and Baysal,
  Olga}]{vadlamani2020studying}
Vadlamani SL, Baysal O.
\newblock Studying Software Developer Expertise and Contributions in Stack
  Overflow and GitHub.
\newblock 2020 36th IEEE International Conference on Software Maintenance and
  Evolution (ICSME), IEEE; 2020. p. 312--323.

\bibitem[{Soliman et~al.(2016)Soliman, Mohamed and Galster, Matthias and
  Salama, Amr R and Riebisch, Matthias}]{soliman2016architectural}
Soliman M, Galster M, Salama AR, Riebisch M.
\newblock Architectural knowledge for technology decisions in developer
  communities: An exploratory study with stackoverflow.
\newblock 2016 13th Working IEEE/IFIP Conference on Software Architecture
  (WICSA), IEEE; 2016. p. 128--133.

\bibitem[{Yin(2017)Yin, Robert K}]{yin2017case}
Yin RK.
\newblock Case study research and applications: Design and methods.
\newblock Sage Publications; 2017.

\bibitem[{Prechelt et~al.(2016)Prechelt, Lutz and Schmeisky, Holger and Zieris,
  Franz}]{prechelt2016quality}
Prechelt L, Schmeisky H, Zieris F.
\newblock Quality experience: a grounded theory of successful agile projects
  without dedicated testers.
\newblock 2016 IEEE/ACM 38th International Conference on Software Engineering
  (ICSE), IEEE; 2016. p. 1017--1027.

\bibitem[{Runeson and H{\"o}st(2009)Runeson, Per and H{\"o}st,
  Martin}]{runeson2009guidelines}
Runeson P, H{\"o}st M.
\newblock Guidelines for conducting and reporting case study research in
  software engineering.
\newblock Empirical Software Engineering 2009;14(2):131.

\bibitem[{Ralph et~al.(2020)Ralph, Paul and Ali, Nauman bin and Baltes,
  Sebastian and Bianculli, Domenico and Diaz, Jessica and Dittrich, Yvonne and
  Ernst, Neil and Felderer, Michael and Feldt, Robert and Filieri, Antonio and
  others}]{ralph2020empirical}
Ralph P, Ali Nb, Baltes S, Bianculli D, Diaz J, Dittrich Y, et~al.
\newblock Empirical Standards for Software Engineering Research.
\newblock arXiv preprint arXiv:201003525 2020;.

\bibitem[{Shahin et~al.(2021)Shahin, Mojtaba and Rezaei-Nasab, Ali and Babar,
  M. Ali}]{onlinedataset}
Shahin M, Rezaei-Nasab A, Babar MA, Interview Questions; 2021.
\newblock \urlprefix\url{https://doi.org/10.5281/zenodo.4935072}.

\bibitem[{Hove and Anda(2005)Hove, Siw Elisabeth and Anda,
  Bente}]{hove2005experiences}
Hove SE, Anda B.
\newblock Experiences from conducting semi-structured interviews in empirical
  software engineering research.
\newblock 2005 11th IEEE International Software Metrics Symposium (METRICS'05),
  IEEE; 2005. p. 10--23.

\bibitem[{Glaser et~al.(1968)Glaser, Barney G and Strauss, Anselm L and
  Strutzel, Elizabeth}]{glaser1968discovery}
Glaser BG, Strauss AL, Strutzel E.
\newblock The discovery of grounded theory; strategies for qualitative
  research.
\newblock Nursing Research 1968;17(4):364.

\bibitem[{Hoda(2011)Hoda, Rashina}]{hoda2011self}
Hoda R.
\newblock Self-organizing agile teams: A grounded theory.
\newblock Victoria University of Wellington 2011;.

\bibitem[{Yang et~al.(2016)Yang, Xin-Li and Lo, David and Xia, Xin and Wan,
  Zhi-Yuan and Sun, Jian-Ling}]{yang2016security}
Yang XL, Lo D, Xia X, Wan ZY, Sun JL.
\newblock What security questions do developers ask? a large-scale study of
  stack overflow posts.
\newblock Journal of Computer Science and Technology 2016;31(5):910--924.

\bibitem[{Rosen and Shihab(2016)Rosen, Christoffer and Shihab,
  Emad}]{rosen2016mobile}
Rosen C, Shihab E.
\newblock What are mobile developers asking about? a large scale study using
  stack overflow.
\newblock Empirical Software Engineering 2016;21(3):1192--1223.

\bibitem[{Hohpe et~al.(2016)Hohpe, Gregor and Ozkaya, Ipek and Zdun, Uwe and
  Zimmermann, Olaf}]{hohpe2016software}
Hohpe G, Ozkaya I, Zdun U, Zimmermann O.
\newblock The software architect's role in the digital age.
\newblock IEEE Software 2016;33(6):30--39.

\bibitem[{Woods(2016)Woods, Eoin}]{woods2016operational}
Woods E.
\newblock Operational: The forgotten architectural view.
\newblock IEEE Software 2016;33(3):20--23.

\bibitem[{Bass(2017)Bass, Len}]{bass2017software}
Bass L.
\newblock The software architect and DevOps.
\newblock IEEE Software 2017;35(1):8--10.

\bibitem[{Haselb{\"o}ck et~al.(2017)Haselb{\"o}ck, Stefan and Weinreich, Rainer
  and Buchgeher, Georg}]{haselbock2017decision}
Haselb{\"o}ck S, Weinreich R, Buchgeher G.
\newblock Decision guidance models for microservices: service discovery and
  fault tolerance.
\newblock Proceedings of the Fifth European Conference on the Engineering of
  Computer-Based Systems, ACM; 2017. p. 1--10.

\bibitem[{Lewis et~al.(2016)Lewis, Grace A and Lago, Patricia and Avgeriou,
  Paris}]{lewis2016decision}
Lewis GA, Lago P, Avgeriou P.
\newblock A decision model for cyber-foraging systems.
\newblock 2016 13th Working IEEE/IFIP Conference on Software Architecture
  (WICSA), IEEE; 2016. p. 51--60.

\bibitem[{Humble and Farley(2010)Humble, Jez and Farley,
  David}]{humble2010continuous}
Humble J, Farley D.
\newblock Continuous delivery: reliable software releases through build, test,
  and deployment automation.
\newblock Pearson Education; 2010.

\bibitem[{Evans(2004)Evans, Eric}]{evans2004domain}
Evans E.
\newblock Domain-driven design: tackling complexity in the heart of software.
\newblock Addison-Wesley Professional; 2004.

\bibitem[{Bosch(2015)Bosch, Jan}]{bosch2015speed}
Bosch J.
\newblock Speed, data, and ecosystems: the future of software engineering.
\newblock IEEE Software 2015;33(1):82--88.

\bibitem[{Cerny et~al.(2017)Cerny, Tomas and Donahoo, Michael J and Pechanec,
  Jiri}]{cerny2017disambiguation}
Cerny T, Donahoo MJ, Pechanec J.
\newblock Disambiguation and comparison of soa, microservices and
  self-contained systems.
\newblock Proceedings of the International Conference on Research in Adaptive
  and Convergent Systems (RACS), ACM; 2017. p. 228--235.

\bibitem[{Gousios et~al.(2016)Gousios, Georgios and Storey, Margaret-Anne and
  Bacchelli, Alberto}]{gousios2016work}
Gousios G, Storey MA, Bacchelli A.
\newblock Work practices and challenges in pull-based development: the
  contributor's perspective.
\newblock 2016 IEEE/ACM 38th International Conference on Software Engineering
  (ICSE), IEEE; 2016. p. 285--296.

\bibitem[{Schauenberg(2014)Schauenberg, Daniel}]{schauenberg2014development}
Schauenberg D, Development, deployment and collaboration at Etsy.
\newblock Web site: [Last accessed: 10/07/2019]; 2014.
\newblock
  \url{https://www.infoq.com/presentations/development-deployment-collaboration-etsy/}.

\bibitem[{Stranghöner(????)R. Stranghöner}]{Self-Contained}
Stranghöner R, Self-Contained Systems. Assembling Software from Independent
  Systems.
\newblock Web site: [Last accessed: 5/2/2019];.
\newblock \url{https://scs-architecture.org/}.

\bibitem[{Dragoni et~al.(2017)Dragoni, Nicola and Giallorenzo, Saverio and
  Lafuente, Alberto Lluch and Mazzara, Manuel and Montesi, Fabrizio and
  Mustafin, Ruslan and Safina, Larisa}]{dragoni2017microservices}
Dragoni N, Giallorenzo S, Lafuente AL, Mazzara M, Montesi F, Mustafin R, et~al.
\newblock Microservices: yesterday, today, and tomorrow.
\newblock Present and Ulterior Software Engineering, Springer; 2017.p.
  195--216.

\bibitem[{Erder and Pureur(2015)Erder, Murat and Pureur,
  Pierre}]{erder2015continuous}
Erder M, Pureur P.
\newblock Continuous architecture: Sustainable architecture in an agile and
  cloud-centric world.
\newblock Morgan Kaufmann; 2015.

\bibitem[{Bergman(2016)G. Bergman}]{Serving}
Bergman G, Serving 86 million users – DevOps the Netflix way.
\newblock Web site: [Last accessed: 25/11/2019]; 2016.
\newblock \url{https://bit.ly/3cLdFFZ}.

\bibitem[{Haff(2017)G. Haff}]{DevOps}
Haff G, DevOps success: A new team model emerges.
\newblock Web site: [Last accessed: 8/8/2019]; 2017.
\newblock \url{https://red.ht/2VTppjx}.

\bibitem[{Guba(1981)Guba, Egon G}]{guba1981criteria}
Guba EG.
\newblock Criteria for assessing the trustworthiness of naturalistic inquiries.
\newblock ECTJ 1981;29(2):75.

\bibitem[{Stol et~al.(2014)Stol, Klaas-Jan and Avgeriou, Paris and Babar,
  Muhammad Ali and Lucas, Yan and Fitzgerald, Brian}]{stol2014key}
Stol KJ, Avgeriou P, Babar MA, Lucas Y, Fitzgerald B.
\newblock Key factors for adopting inner source.
\newblock ACM Transactions on Software Engineering and Methodology (TOSEM)
  2014;23(2):1--35.

\bibitem[{Wiggins(????)A. Wiggins}]{Twelve-Factor}
Wiggins A, Twelve-Factor App methodology.
\newblock Web site: [Last accessed: 10/9/2019];.
\newblock \url{https://12factor.net/}.

\bibitem[{Cito et~al.(2015)Cito, J{\"u}rgen and Leitner, Philipp and Fritz,
  Thomas and Gall, Harald C}]{cito2015making}
Cito J, Leitner P, Fritz T, Gall HC.
\newblock The making of cloud applications: An empirical study on software
  development for the cloud.
\newblock Proceedings of the 2015 10th Joint Meeting on Foundations of Software
  Engineering (ESEC/FSE), ACM; 2015. p. 393--403.

\bibitem[{T{\'o}masd{\'o}ttir et~al.(2017)T{\'o}masd{\'o}ttir, Krist{\'\i}n
  Fj{\'o}la and Aniche, Mauricio and van Deursen, Arie}]{tomasdottir2017and}
T{\'o}masd{\'o}ttir KF, Aniche M, van Deursen A.
\newblock Why and how JavaScript developers use linters.
\newblock 2017 32nd IEEE/ACM International Conference on Automated Software
  Engineering (ASE), IEEE; 2017. p. 578--589.

\end{thebibliography}

\end{document}